\documentclass[journal,draftcls,onecolumn,12pt,twoside]{IEEEtran}
\usepackage{graphicx}
\usepackage{caption}
\newcommand{\squeezeup}{\vspace{-2.5mm}}

\usepackage{hyperref}
\usepackage{bbm}
\usepackage{bm}

\usepackage{pgfplots}
\usepackage[utf8]{inputenc}
\usepackage[english]{babel}

\usepackage{amssymb}   
\usepackage{stfloats}

\usepackage{mathtools}

\usepackage{accents}

 \usepackage{booktabs}

\DeclareMathOperator*{\argmax}{arg\,max}

\newcommand\munderbar[1]{%
  \underaccent{\bar}{#1}}

 \newcommand{\Qu}[2][i]{\ensuremath{Q_{#1,#2}}} 
  \newcommand{\Es}[2][i]{\ensuremath{E_{s,#1#2}}} 
    \newcommand{\EsCR}[2][i]{\ensuremath{E_{s,#1#2}}}

\newcommand{\Larezou}[2][i]{\ensuremath{\lambda_{#2#1}}}

\newcommand{\Au}[2][i]{\ensuremath{\mathcal A_{#1}^{#2}}} 
\newcommand{\Bu}[2][i]{\ensuremath{\mathcal B_{#1}^{#2}}(\gamma_{#1})}

\newcommand{\Cua}[2][i]{\ensuremath{\mathcal B_{#1}^{#2}}}

\usepackage{colortbl}

\newtheorem{theorem}{Theorem}

\newtheorem{lemma}[theorem]{Lemma}
\newtheorem{proposition}{Proposition}

\pgfplotsset{width=5cm,compat=newest,every axis legend/.append style={at={(0.5,1.35)}, anchor=south}}

\begin{document}
\title{Joint Source-Channel Coding for the Multiple-Access Channel with Correlated Sources}

\author{Arezou Rezazadeh, Josep Font-Segura, Alfonso Martinez,\\ and Albert Guill{\'e}n i F{\`a}bregas
\thanks{A. Rezazadeh, J. Font-Segura and A. Martinez  are with the Department of Information and Communication Technologies, Universitat Pompeu Fabra, Barcelona 08018, Spain (e-mail: arezou.rezazadeh@upf.edu, \{josep.font-segura, alfonso.martinez\}@ieee.org).}
\thanks{A. Guill{\'e}n i F{\`a}bregas is with Instituci{\'o} Catalana de Recerca i Estudis Avan\c{c}ats, the Department of Information and Communication Technologies, Universitat Pompeu Fabra, Barcelona 08018, Spain, and also with the Department of Engineering, University of Cambridge, Cambridge CB2 1PZ, U.K. (e-mail: guillen@ieee.org).}
\thanks{This work has been funded in part by the European Research Council under grant 725411, and by the Spanish Ministry of Economy and Competitiveness under grant TEC2016-78434-C3-1-R.}
}

\maketitle
\thispagestyle{empty}
\begin{abstract}
This paper studies
the random-coding exponent of joint source-channel coding for
the multiple-access channel with correlated sources.
For each user, by defining a threshold,
 the messages of each source are partitioned into two classes.
The achievable exponent for correlated sources
with two message-dependent input distributions for each user is determined and 
  shown to be larger than that achieved using
only one input distribution for each user.
A system of equations is presented to determine the
optimal thresholds 
maximizing the achievable exponent. The obtained exponent is compared with the 
one derived for the MAC with independent sources.

\end{abstract} 
\IEEEpeerreviewmaketitle



\section{Introduction}
\label{part1_isit2019}
Some studies show that for point-to-point communication, 
using 
 a partition of the message set into source-type classes
and assigning one input distribution for each class leads to a larger
exponent than having codewords drawn from a single product distribution \cite{Cs2,jscc}.
Recent studies generalize this result to the multiple-access channel (MAC) \cite{farkas}. 
In \cite{itw2018}, the exponent with message-dependent random-coding across two classes is found to beat independent identically distributed (iid) random-coding  for a
 two-user MAC with independent sources.
 %
 
For a two-user MAC with correlated sources, \cite{isit2017} studied a message-dependent ensemble where
codewords are generated by a symbol-wise conditional probability
distribution that depends on the
instantaneous source symbol and on the empirical distribution of the source
sequence.
The derived exponent 
were given 
 as a multidimensional
optimization problem over distributions  
 i.e. primal domain  
\cite{isit2017}.
In this paper, 
we apply  Lagrange duality theory to the results in \cite{isit2017} and find the exponent in the dual domain, i.e. as
a lower dimensional problem over parameters in terms of Gallager functions.
We show that the obtained exponent is larger than
that achieved using only one input distribution for each user.

\subsection{System Model, Definitions and Notations}
Using the 
convention that scalar random variables are denoted by capital letters, we consider two correlated sources characterized by
$P_{U_1U_2}\in \mathcal P_{\mathcal U_1 \mathcal U_2}$ 
on the alphabet $\mathcal{U}_1\times\mathcal{U}_2$, where $\mathcal{U}_1$ and $\mathcal{U}_2$ are the respective source alphabets, and $\mathcal P_{\mathcal U_1 \mathcal U_2}$ is the set of all  possible distributions of  $(U_1,U_2)$.
In addition, 
the set of all empirical distributions on a joint
vector in $\mathcal{U}_1^n\times\mathcal{U}_2^n$
(i.e. types) is denoted by $\mathcal P_{\mathcal U_1 \mathcal U_2}^n$. 
Given $\hat P_{U_1U_2}\in \mathcal P_{\mathcal U_1 \mathcal U_2}^n$, the joint-type class 
$\mathcal T^n(\hat P_{U_1U_2})$ is the set of all joint sequences in $\mathcal{U}_1^n\times\mathcal{U}_2^n$ with joint type $\hat P_{U_1U_2}$.

Encoder $\nu=1,2$ maps a length-$n$
source message $\bm u_\nu$ to the length-$n$ codeword $\bm x_\nu(\bm u_\nu)$ drawn from the codebook $\mathcal{C}^\nu=\{\bm x_\nu(\bm u_\nu); \bm u_\nu\in \mathcal{U}_\nu^n\}$. 
To simplify
some expressions, we use underline  to represent a pair of quantities for users 1 and 2, such as $\munderbar {u}= (u_1,u_2)$, $\munderbar{\bm u} = (\bm u_1,\bm u_2)$ or $\munderbar{\mathcal U}=\mathcal U_1\times \mathcal U_2$.  
Both users send their respective codewords  over  discrete memoryless MAC with transition probability $W(y|\munderbar x)$, input alphabets $\mathcal{X}_1$ and $\mathcal{X}_2$, 
 and output alphabet $\mathcal{Y}$.
By receiving the sequence $\bm y$, the decoder estimates the transmitted pair messages $\munderbar{\bm u}$ based on the maximum a posteriori criterion:
\begin{align}
\hat{\munderbar{\bm u}}=\argmax_{\munderbar{\bm u}\in \munderbar{\mathcal{U}}^n}P^n_{\munderbar{\bm U}}(\munderbar{\bm u})W^n\big(\bm y\,|\,\bm x_1(\bm u_1),\bm x_2(\bm u_2)\big).\label{criteria}
\end{align}

An error occurs if the decoded messages $\hat{\munderbar{\bm u}}$ differ from the transmitted $\munderbar{\bm u}$; the error
probability for a given pair of codebooks is thus given by
\begin{align}
\epsilon^n(\mathcal{C}^1,\mathcal{C}^2)\triangleq {\mathbb P}\left[\hat{\munderbar{\bm U}}\neq \munderbar{\bm U}\right].
\end{align}
The error event $\hat{\munderbar{\bm U}}\neq \munderbar{\bm U}$ can be split into three disjoint types of error events, namely  $(\hat{ \bm U}_1, {\bm U}_2)\neq (\bm U_1, \bm U_2)$, $({ \bm U}_1, \hat{\bm U}_2)\neq (\bm U_1, \bm U_2)$ and $(\hat{ \bm U}_1, \hat{\bm U}_2)\neq (\bm U_1, \bm U_2) $. These events are respectively labelled by $\tau$, with
$\tau\in\{\{1\},\{2\},\{1,2\}\}$. 
To further simplify some expressions, we adopt the following convention,
\begin{align}
u_{\tau}=\left\{
\begin{array}{ll}
\emptyset  & \tau=\emptyset\\
u_1  & \tau=\{1\}\\
u_2   & \tau=\{2\}\\
\munderbar u  & \tau=\{1,2\}
\end{array} 
\right.,
\end{align}
for the variable $u_\nu$, and similarly for the probability distribution $Q_\nu$ and the set $\mathcal X_\nu$. We denote the complement of $\nu$ (or $\tau$) in the set $\{1,2\}$ (or the subsets of $\{1,2\}$) by $\nu^c$ (or $\tau^c$), e.g. $\tau^c=\{2\}$ for $\tau=\{1\}$ and $\tau^c=\emptyset$ for $\tau=\{1,2\}$. 

The pair of sources $(U_1,U_2)$ is transmissible over the channel if there exists a sequence of codebooks $(\mathcal{C}^1_n,\mathcal{C}^2_n)$ such that $\lim_{n\rightarrow\infty} \epsilon^n(\mathcal{C}^1_n,\mathcal{C}^2_n)=0$. An exponent $E(P_{\munderbar{U}},W)$ is achievable if there exists a sequence of codebooks such that
\begin{align}
 \liminf_{n\rightarrow \infty} -\dfrac{1}{n} \log\left (\epsilon^n(\mathcal{C}^1_n,\mathcal{C}^2_n)\right )\geq E(P_{\munderbar{U}},W).
\end{align}

\section{Joint Source-Channel Random-Coding}
For point-to-point transmission of a discrete memoryless
source $P_U$
over a discrete memoryless channel $W$, the joint source-channel iid random-coding
 with input distribution $Q$ is expressed in terms of Gallager source and channel functions, respectively given by \cite{gala}
\begin{align}
    &E_s(\rho,P_U)=\left(1+\rho\right)\log\left(\sum_u P_U(u)^{\frac{1}{1+\rho}}\right),
    \label{Es_isit2019}
\\
    &E_0(\rho,Q,W)=-\log
\sum_{y} \left(\sum_{x}
 Q(x)
W(y|x)^{\frac{1}{1+\rho}}
\right)^{1+\rho} 
.\label{E0_tau_isit2019}
\end{align}

For a two-user MAC with  two correlated sources $P_{\munderbar U}$, transition probability $W$ and given input distributions $Q_1$ and $Q_2$,  the i.i.d random-coding  exponent
is given by
\begin{align}
E^{\text{i.i.d}}(P_{\munderbar U},W)=
\min_{\tau\in\left\{
\{1\},\{2\},\{1,2\}\right\}}
\max_{\rho\in[0,1]}E_0(\rho,Q_{\tau},WQ_{\tau^c})-E_{s,\tau}(\rho,P_{\munderbar U}),
\label{2019.04.17_21.57}
\end{align}
where  $E_0(\cdot)$
is given by \eqref{E0_tau_isit2019}, and
$E_{s,\tau}(\cdot)$ is the
 generalized Gallager's source  functions for error type $\tau$ where
 \setlength{\arraycolsep}{0.0em}
\begin{align}
&E_{s,\tau}(\rho,P_{\munderbar U})=
\log
\sum_{u_{\tau^c}}\left(\sum_{u_\tau}
P_{ \munderbar U}(\munderbar u)^{\frac{1}{1+\rho}} 
\right)^{1+\rho} 
.\label{Es_tau_isit2019}
\end{align}

We recall that in \eqref{2019.04.17_21.57},
for  $\tau=\{1\}$ and $\tau=\{2\}$,  $WQ_{\tau^c}$  denotes 
 a point-to-point channel  
 with input and output alphabets given by $\mathcal X_\tau$ and  $\mathcal X_{\tau^c}\times \mathcal Y$, respectively. For $\tau=\{1,2\}$, 
 the input distribution $Q_\tau$ is the product distribution $Q_1(x_1)Q_2(x_2)$ over the alphabet $\mathcal X_1 \times \mathcal X_2$, and $WQ_{\tau^c}=W$.

Another possible strategy, known as message-dependent random-coding \cite{itw2018},
is to assign source messages to disjoint classes, and to use codewords
generated according to a distribution that depends on the class index.
The primal form of 
the message-dependent  exponent 
for the MAC with two correlated sources has been given by \cite[Eq. (9)]{isit2017} where codewords are generated 
according to  conditional input distributions
that depend on the composition of the source message. 

For user $\nu=1,2$, let the set of input distributions 
$\big\{\Qu[\nu]{\hat P_{U_\nu}} \big\}$ be given.
By applying the same approach 
in  \cite{isit2017},
 the achievable exponent of \cite[Eq. (9)]{isit2017} 
for statistically independent messages and  codewords  is simplified to
\begin{align}
&E_1(P_{\munderbar U},W)=\min_{\tau\in\{\{1\},\{2\},\{1,2\}\}}
\min_{\hat{P}_{\munderbar U} \in\mathcal{P}_{ \mathcal{\munderbar U}} }\min_{\hat{P}_{\munderbar  X Y}\in\mathcal{P}_{ \mathcal{\munderbar X}\times\mathcal{Y}}}
D(\hat P_{\munderbar U }|| P_{\munderbar U})
+
D(\hat P_{\munderbar  X Y}||Q_{1,\hat P_{U_1}}Q_{2,\hat P_{U_2}}W)
\nonumber\\
&\hspace{1em}+ \left[
\min_{\substack{ \tilde{P}_{\munderbar U }
\in \mathcal{P}_{\mathcal{\munderbar U}} :
\tilde P_{ U_{\tau^c}}  =\hat P_{ U_{\tau^c}  },\ \\ \mathbb{E}_{\tilde{P}}\log P_{\munderbar U} \geq \mathbb{E}_{\hat P}\log P_{\munderbar U}
}}
\min_{\substack{\tilde{P}_{ \munderbar X Y}\in \mathcal{P}_{ \mathcal{\munderbar X}\times\mathcal{Y}} : 
\tilde P_{   X_{\tau^c} Y}=\hat P_{   X_{\tau^c} Y},\\ \mathbb{E}_{\tilde{P}}\log W \geq \mathbb{E}_{\hat P}\log W
}}
D(\tilde{P}_{\munderbar  X Y}|| Q_{\tau,\tilde P_{U_\tau}}\hat{P}_{X_{\tau^c}Y})
-H(\tilde P_{U_\tau|U_{\tau^c}})
\right]^+ \label{primal_exp_isit2019}
,\end{align}
where $[x]^+=\max\{0,x\}$.
In order to
 find the dual-domain version of \eqref{primal_exp_isit2019}, we
 firstly  analyze the source-exponent terms.


\subsection{Source exponent function}
\label{sub_source_exponent_func_isit2019}
In  \cite{itw2018}, for each user, a fixed threshold
 was  considered to partition the source-message set  into two classes, i.e.,
\begin{align}
\Au[\nu]{1}(\gamma_\nu)=\big\{\bm u_\nu\in \mathcal U^n_\nu \,:\, P_{\bm U_\nu}^n(\bm u_\nu)\geq \gamma_\nu^n\big\}\label{A1_itw2018},\\
\Au[\nu]{2}(\gamma_\nu)=\big\{\bm u_\nu\in \mathcal U^n_\nu \,:\, P_{\bm U_\nu}^n(\bm u_\nu)< \gamma_\nu^n\big\}.
\label{A2_itw2018}
\end{align}
  Here, we use the same idea in the primal domain.  Exploiting that  the source messages are encoded independently for each user in distributed source coding \cite{SWsource}, 
  the following Lemma gives
the asymptotic form of  
\eqref{A1_itw2018} and   \eqref{A2_itw2018} for correlated sources.
\begin{lemma}
\label{mac_partition_isit2019}
Let $P_{\munderbar U}$ be a probability distribution of two correlated sources and 
 $P_{U_\nu}$ be the marginal distribution for source $\nu=1,2$. Given partitioning thresholds $\gamma_\nu\in[0,1]$,
the set of probability distributions
$ \mathcal{P}_{\munderbar{\mathcal{U}}}$ 
can be partitioned into disjoint classes $\Bu[\nu]{1}$ and $\Bu[\nu]{2}$ where
\begin{align}
\Bu[\nu]{1}=\left\{ 
\hat P_{\munderbar U}\in \mathcal{P}_{\munderbar{\mathcal {U}}}
:\sum_{\munderbar u}\hat P_{\munderbar U} ( \munderbar u) \log P_{U_\nu}(u_\nu) \geq \log(\gamma_\nu)
\right\}\label{B1_isit2019},\\
\Bu[\nu]{2}=\left\{ 
\hat P_{\munderbar U}\in \mathcal{P}_{\munderbar{\mathcal {U}}}
:\sum_{\munderbar u}\hat P_{\munderbar U} ( \munderbar u) \log P_{U_\nu}(u_\nu)<\log(\gamma_\nu)
\right\}.
\label{B2_isit2019}
\end{align}
\end{lemma}

\begin{IEEEproof}
See Appendix \ref{Proof_mac_partition}.
\end{IEEEproof}
Roughly speaking, $\Bu[\nu]{1}$   in \eqref{B1_isit2019}, can be interpreted as the
asymptotic limit of the union  of
sequences $(\bm u_1,\bm u_2)$ with type  $\hat P^n_{\munderbar U}$, 
where as long as the marginal probability $P_{U_\nu}^n(\bm u_\nu)$ is not less  than the threshold $\gamma_\nu^n$,
the empirical distribution of 
$\bm u_{\nu^c}$ can be arbitrary (similarly for $\Bu[\nu]{2}$   in \eqref{B2_isit2019}).
The following Proposition finds the Gallager
source exponent function for the messages corresponding to 
$\Bu[\nu]{1}$  and $\Bu[\nu]{2}$.

\begin{proposition}
\label{dif_Esi_isit2019}
For
given  $\gamma_\nu\in [0,1]$ and 
$i_\nu\in\{1,2\}$,
in view of $\Bu[\nu]{i_\nu}$  given by \eqref{B1_isit2019} and \eqref{B2_isit2019}, we have
\setlength{\arraycolsep}{0.0em}
\begin{align}
\min_{\substack{\hat P_{\munderbar U}\in \mathcal P_{\munderbar{\mathcal U}}:\hat P_{\munderbar U}\in\Bu[1]{i_1},\hat P_{\munderbar U}\in\Bu[2]{i_2}}}D(\hat P_{\munderbar U}||P_{\munderbar U})
-\rho H(\hat P_{U_\tau|U_{\tau^c}})
=
-\EsCR[\tau,i_1,i_2]{}(\rho,P_{\munderbar U},\munderbar \gamma),
\label{bef_6_isit2019}
\end{align}
where 
\begin{align}
\EsCR[\tau,i_1,i_2]{}(\rho,P_{\munderbar U},\munderbar \gamma)=
\hspace{27em}\nonumber\\
\min_{\Larezou[1]{}\geq 0,\Larezou[2]{}\geq 0}
\log \sum_{u_{\tau^c}} \Bigg(
\sum_{ u_\tau}P_{\munderbar U}(\munderbar u)^{\frac{1}{1+\rho}}
\left(\frac{P_{U_1}(u_1)}{\gamma_1}
\right)
^{-\frac{(-1)^{i_1}\Larezou[1]{}}{1+\rho}}
\left(\frac{P_{U_2}(u_2)}{\gamma_2}
\right)^{-\frac{(-1)^{i_2}\Larezou[2]{}}{1+\rho}}
\Bigg)^{1+\rho}.\label{6.1_isit2019}
\end{align}
\end{proposition}
\begin{IEEEproof} 
 See Appendix \ref{Proof_dif_Esi_isit2019}.
 \end{IEEEproof} 
In fact, in \eqref{6.1_isit2019}, the objective function is a convex  function with respect to $\Larezou[\nu]{}$ for $\nu=1,2$, and the optimal $\Larezou[\nu]{}$ minimizing  \eqref{6.1_isit2019} are the solution of an implicit equation  obtained 
by setting
the partial derivative of the objective function of  \eqref{6.1_isit2019} with respect to $\Larezou[\nu]{}$
equal to zero.
 To be precise, 
 for the cases where both constraints $\hat P_{\munderbar U}\in\Bu[1]{i_1}$ and $\hat P_{\munderbar U}\in\Bu[2]{i_2}$  are active,  $\Larezou[1]{}$
and $\Larezou[2]{}$ derived as the solution of the 
implicit equation, 
are greater than zero. Otherwise, the 
solution of the implicit equation is negative and the
optimal 
$\Larezou[\nu]{}$ is zero.

Here, we  compare the result given by \eqref{6.1_isit2019} with that for independent sources. In \cite{jscc,itw2018}, it was shown that the exponent is expressed in terms of two
%
$\Es[i_\nu]{}(\cdot)$ functions, namely 
\setlength{\arraycolsep}{0.0em}
\begin{align}
\Es[1]{}(\rho,P_{U_\nu},\gamma_\nu)=
\left\{
\begin{array}{rl}
&E_s(\rho,P_{U_\nu})\hspace{11.5em} \frac{1}{1+\rho} \geq\frac{1}{1+\rho_{\gamma_\nu}},
\\
&E_s(\rho_{\gamma_\nu},P_{U_\nu})+
E_s^\prime(\rho_{\gamma_\nu})(\rho-\rho_{\gamma_\nu})
\hspace{2.5em} 
\frac{1}{1+\rho} <\frac{1}{1+\rho_{\gamma_\nu}},
\end{array} \right.\label{Esi_itw_2018_1}
\end{align}
and a similar definition for $\Es[2]{}(\rho,P_{U_\nu},\gamma_\nu)$, with the two conditions swapped.
In the definition of the $\Es[i_\nu]{}(\cdot)$ functions,
the parameter $\rho_{\gamma_\nu}$ is the solution of the implicit equation
\begin{align}
\frac{\sum_u P_{U_\nu}(u_\nu)^{\frac{1}{1+\rho}}\log P_{U_\nu}(u_\nu)}{\sum_{u_\nu} P_{U_\nu}(u_\nu)^{\frac{1}{1+\rho}}}=\log(\gamma_\nu),
\label{11}
\end{align}
as long as $\min_{u_\nu} P_{U_\nu}(u_\nu) \leq\gamma_\nu\leq \max_{u_\nu} P_{U_\nu}(u_\nu)$. 
If $\gamma_\nu\in[0,\min_{u_\nu} P_{U_\nu}(u_\nu))$, we have $\rho_{\gamma_\nu}=-1_-$ and 
if $\gamma_\nu\in(\max_{u_\nu} P_{U_\nu}(u_\nu),1]$, we have $\rho_{\gamma_\nu}=-1_+$. 

 Additionally,
from  \cite[Lemma 3]{arzou_arxiv}, 
 for each  source $\nu=1,2$ with distribution $P_{U_\nu}$, 
 threshold $\gamma_\nu$, 
 and $i_\nu=1,2$,
 we have
 \begin{align}
   \Es[i_\nu]{}(\rho,P_{U_\nu},\gamma_\nu) =
   \min_{\Larezou[\nu]{}\geq 0} \log \sum_{u_\nu} 
   P_{ U_\nu}( u_\nu)^{\frac{1}{1+\rho}}
  \left(\frac{P_{U_\nu}(u_\nu)}{\gamma_\nu}
\right)
^{-\frac{(-1)^{i_\nu}\Larezou[\nu]{}}{1+\rho}}.
\label{mik_beram_isit2019}
 \end{align}

For independent sources, by applying $P_{\munderbar U}(\munderbar u)=P_{U_1}(u_1)P_{U_2}(u_2)$ in \eqref{6.1_isit2019},  and in view of \eqref{mik_beram_isit2019}, 
the function $\EsCR[\tau,i_1,i_2]{}(\rho,P_{\munderbar U},\munderbar \gamma)$ is simplified as
\begin{align}
    \EsCR[\tau,i_1,i_2]{}(\rho,P_{U_1}(u_1)P_{U_2}(u_2),
    \munderbar \gamma)=
    \Es[i_\tau]{}(\rho,P_{U_\tau},\gamma_\tau)
    +\Es[i_{\tau^c}]{}(0,P_{U_{\tau^c}},\gamma_{\tau^c}),
    \label{Esi1i2_indep_sources_isit2019}
\end{align}
where  as discussed 
in  \cite[Eq. (15)]{itw2018}, for $\tau=\{1,2\}$, $\Es[i_{\{1,2\}}]{}(\rho,P_{\munderbar U},\munderbar \gamma)=
\Es[i_1]{}(\rho,P_{U_1},\gamma_1)+\Es[i_2]{}(\rho,P_{U_2},\gamma_2)$. In fact, depending on the tangent points in \eqref{11}, 
$\EsCR[\{1,2\},i_1,i_2]{}(\cdot)$ as a function of $\rho$ is either
$ E_s(\rho,P_{U_\nu})+E_s(\rho,P_{U_{\nu^c}})$
or
$ E_s(\rho,P_{U_\nu})+\Es[i_{\nu^c}]{}(\rho,P_{U_{\nu^c}},\gamma_{\nu^c})$ where 
$\nu$ can be $1$ or $2$, and $\nu^c$
denotes the complement index of $\nu$
among the set $\{1,2\}$.

\begin{figure}[!t]
\centering
\footnotesize
\input{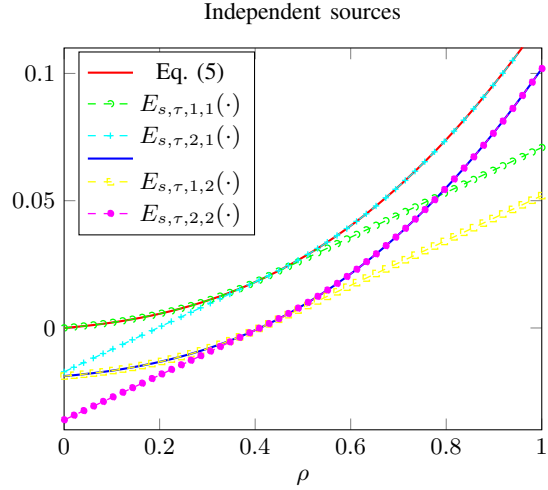}
\caption{
$\EsCR[\tau,i_1,i_2]{}(\cdot)$ in \eqref{Esi1i2_indep_sources_isit2019} for  two independent sources
 versus $\rho$,
for fixed  $\gamma_1$ and $\gamma_2$ where $i_1,i_2=1,2$.
For error type $\tau\in\left\{\{1\},\{2\}\right\}$, the solid red and blue curves are respectively $E_s(\rho,P_{U_\tau})$ and 
$E_{s}(\rho,P_{U_\tau})+\Es[i_{\tau^c}]{}(0,P_{U_{\tau^c}},\gamma_{\tau^c})$. }
\squeezeup
\label{fig_ind}
\end{figure}

For  error type $\tau\in\left\{\{1\},\{2\}\right\}$ and for the four combinations of $i_1,i_2\in\{1,2\}$,
Fig. \ref{fig_ind} shows \eqref{Esi1i2_indep_sources_isit2019} for two independent sources with given $\gamma_1$, $\gamma_2$. As  shown in \eqref{Esi1i2_indep_sources_isit2019} and for 
Fig. \ref{fig_ind}, the functions $ \EsCR[\tau,1,1]{}(\cdot)$ and $ \EsCR[\tau,2,1]{}(\cdot)$ follow $E_{s}(\rho,P_{U_\tau})$ given by \eqref{Es_isit2019}, for an interval of $\rho$, while they are the straight line tangent to Gallager's source  function beyond that
interval.
However, 
the functions $ \EsCR[\tau,1,2]{}(\cdot)$ and $ \EsCR[\tau,2,2]{}(\cdot)$  are either
the Gallager's source  function shifted by $\Es[i_{\tau^c}]{}(0,P_{U_{\tau^c}},\gamma_{\tau^c})$
or the straight line tangent to it.

\begin{figure}[!t]
\centering
\footnotesize
\input{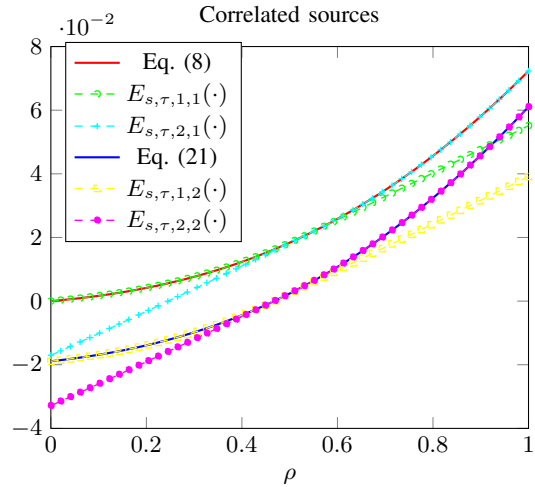}
\caption{
$\EsCR[\tau,i_1,i_2]{}(\cdot)$ in \eqref{6.1_isit2019} for  two correlated sources
 versus $\rho$,
for fixed  $\gamma_1$ and $\gamma_2$ where $i_1,i_2=1,2$.
The solid red and blue curves are respectively given by \eqref{Es_tau_isit2019} and \eqref{envelop_isit2019}. }
\squeezeup
\label{fig_cor}
\end{figure}

On the other hand, 
for correlated sources with 
 four combinations of $i_1,i_2\in\{1,2\}$, 
Fig. \ref{fig_cor} shows \eqref{6.1_isit2019} for two correlated sources with given $\gamma_1$, $\gamma_2$ and  error type $\tau$. It can be seen that 
for the example of 
Fig. \ref{fig_cor}, the functions $ \EsCR[\tau,1,1]{}(\cdot)$ and $ \EsCR[\tau,2,1]{}(\cdot)$ are the generalized Gallager's source function \eqref{Es_tau_isit2019}  for an interval of $\rho$, while they are a curve tangent to 
$E_{ s,\tau}(\cdot)$
beyond that
interval.
Thus, unlike the independent sources,   instead of a straight line tangent to 
Gallager's source function, 
for correlated sources, a
curve is tangent to 
$E_{ s,\tau}(\cdot)$. 
The reason for this is explained in the following.

In Fig. \ref{fig_cor},
consider $ \EsCR[\tau,2,1]{}(\cdot)$ where $i_1=2$ and 
$i_2=1$. For the region of $\rho$ where $ \EsCR[\tau,2,1]{}(\cdot)$ equals to 
$E_{s,\tau}(\cdot)$, both constraints
$\hat P_{\munderbar U}\in\Bu[1]{2}$ and
$\hat P_{\munderbar U}\in\Bu[2]{1}$
are inactive, while for the region of 
$\rho$ where $ \EsCR[\tau,2,1]{}(\cdot)$ equals to the curve  tangent to 
$E_{ s,\tau}(\cdot)$, only one of  the  constraints
$\hat P_{\munderbar U}\in\Bu[1]{2}$ or
$\hat P_{\munderbar U}\in\Bu[2]{1}$ is active (similarly for $ \EsCR[\tau,1,1]{}(\cdot)$).
 For given $i_1,i_2$, let $\nu\in\{1,2\}$ correspond to the  active constraint.
For example, in Fig. \ref{fig_cor}, for the region of 
$\rho$ where $ \EsCR[\tau,2,1]{}(\cdot)$ equals the tangent curve,  only the constraint $\hat P_{\munderbar U}\in\Bu[\nu]{i_\nu}$ is active. Then,
the primal form of the curve is 
\begin{align}
-\min_{\substack{
\hat P_{\munderbar U}\in \mathcal{P}_{\munderbar{\mathcal {U}}}
:\\\sum_{\munderbar u}\hat P_{\munderbar U} ( \munderbar u) \log P_{U_\nu}(u_\nu) = \log(\gamma_\nu)
}}D(\hat P_{\munderbar U}||P_{\munderbar U})
-\rho H(\hat P_{U_\tau|U_{\tau^c}}),
\label{curve_primal_isit2019}
\end{align}
as
corresponds to  the Gallager's source 
exponent function of messages source $\nu$ whose empirical distributions are fixed, i.e.,  $\big\{
\hat P_{\munderbar U}\in \mathcal{P}_{\munderbar{\mathcal {U}}}
:\sum_{\munderbar u}\hat P_{\munderbar U} ( \munderbar u) \log P_{U_\nu}(u_\nu) = \log(\gamma_\nu)
\big\}$. 


We note that \eqref{curve_primal_isit2019} describes the situation where only the type class of one source is fixed. Thus, we have more freedom in the source type class of the other source. This implies that for correlated sources 
 the joint type class is not fixed, but rather contains the union of joint type classes whose type class of one of the sources is fixed.
Unlike for independent sources, for correlated sources
\eqref{curve_primal_isit2019} is a curve rather than a straight line.

Coming back to Fig. \ref{fig_cor}, for an interval of $\rho$,
the function $ \EsCR[\tau,1,2]{}(\cdot)$ ($ \EsCR[\tau,2,2]{}(\cdot)$) is
\begin{align}
    \min_{\lambda_\nu \geq 0}
    \log \sum_{u_{\tau^c}} \Bigg(
\sum_{ u_\tau}P_{\munderbar U}(\munderbar u)^{\frac{1}{1+\rho}}
\left(\frac{P_{U_\nu}(u_\nu)}{\gamma_\nu}
\right)
^{-\frac{(-1)^{i_\nu}\Larezou[\nu]{}}{1+\rho}}
\Bigg)^{1+\rho},
\label{envelop_isit2019}
\end{align}
where $\nu\in\{1,2\}$ indicates that only the constraint $\hat P_{\munderbar U}\in\Bu[\nu]{i_\nu}$ is active. In addition, beyond that
interval of $\rho$, 
the functions $ \EsCR[\tau,1,2]{}(\cdot)$ ($ \EsCR[\tau,2,2]{}(\cdot)$) is \eqref{6.1_isit2019} where both constraints $\hat P_{\munderbar U}\in\Bu[1]{i_1}$ and
$\hat P_{\munderbar U}\in\Bu[2]{i_2}$ are active.

\subsection{Error Exponent Analysis}
The primal form of 
the message-dependent  exponent 
for the MAC with two correlated sources is given by \eqref{primal_exp_isit2019}.
To find the dual-domain of 
\eqref{primal_exp_isit2019},
 we use the following Lemma.
\begin{lemma}
\label{remove_one_mini_isit2019}
$E_1(P_{\munderbar U},W)$ given by \eqref{primal_exp_isit2019} is bounded  as
\begin{align}
E_1(P_{\munderbar U},W)\geq
 \min_{\tau}
\min_{\hat{P}_{\munderbar U}
 \in\mathcal{P}_{ \mathcal{\munderbar U}}
 }\min_{\hat{P}_{\munderbar  X Y}
 \in\mathcal{P}_{ \mathcal{\munderbar X}\times\mathcal{Y}}
 } D(\hat P_{\munderbar U }|| P_{\munderbar U})
+D(\hat P_{\munderbar  X Y}||Q_{1,\hat P_{U_1}}Q_{2,\hat P_{U_2}}W)
\hspace{5em}\nonumber\\
+\max_{\rho\in[0,1]}
 \rho D(\hat{P}_{ \munderbar X Y}||\Qu[\tau]{\hat P_{U_\tau}}\hat P_{X_{\tau^c}Y})
-\rho H(\hat P_{U_\tau|U_{\tau^c}})
.\label{nef_alf_15_06_29_10_2018}
\end{align}
\end{lemma}
\begin{IEEEproof}
See Appendix \ref{proof_remove_one_mini_isit2019}.
\end{IEEEproof}

The optimization problem over $\hat P_{\munderbar X,Y}$  in \eqref{nef_alf_15_06_29_10_2018} is coupled with the minimization problem over 
$\hat P_{\munderbar U}$ through $\Qu[\nu]{\hat P_{U_\nu}}$ for $\nu=1,2$.
In view of classes defined by \eqref{B1_isit2019} and \eqref{B2_isit2019}, 
 we express the dependency of the 
 input distribution $\Qu[\nu]{\hat P_{U_\nu}}$ on $\hat P_{U_\nu}$, through  the class index.
In other words, for  $\hat P_{U_\nu}\in \Bu[\nu]{1}$, we let  $\Qu[\nu]{\hat P_{U_\nu}}=\Qu[\nu]{1}$ and similarly 
for  $\hat P_{U_\nu}\in \Bu[\nu]{2}$, we let  $\Qu[\nu]{\hat P_{U_\nu}}=\Qu[\nu]{2}$. Applying this  to \eqref{nef_alf_15_06_29_10_2018}, and 
splitting the minimization over $\hat P_{\munderbar U}$ into minimization over disjoint classes as 
$\min_{i_1,i_2=1,2} \min_{\substack{\hat P_{\munderbar U}\in \mathcal P_{\munderbar{\mathcal U}}:\hat P_{\munderbar U}\in\Bu[1]{i_1}, \hat P_{\munderbar U}\in\Bu[2]{i_2}}}$,
we find that
\begin{align}
E_1(P_{\munderbar U},W)\geq
 \min_{\tau} \min_{i_1,i_2=1,2}
 \min_{\substack{\hat P_{\munderbar U}\in \mathcal P_{\munderbar{\mathcal U}}:\hat P_{\munderbar U}\in\Bu[1]{i_1}, \hat P_{\munderbar U}\in\Bu[2]{i_2}}}D(\hat P_{\munderbar U}||P_{\munderbar U})
 \hspace{10em}\nonumber\\
+
\min_{\hat{P}_{\munderbar X Y}\in\mathcal{P}_{\mathcal{\munderbar X}\times\mathcal{Y}}
}
D(\hat P_{\munderbar X Y}|| \Qu[1]{i_1}\Qu[2]{i_2}W)
+ \max_{\rho\in[0,1]}\rho D(\hat{P}_{ \munderbar X Y}||\Qu[\tau]{i_\tau}\hat P_{X_{\tau^c}Y})-\rho H(\hat P_{U_\tau|U_{\tau^c}}).\label{2_isit2019}
\end{align} 

By using the min-max inequality,
  we swap the maximization over $\rho$ with the minimizations over $\hat{P}_{\munderbar X Y}\in\mathcal{P}_{\mathcal{\munderbar X}\times\mathcal{Y}}$ and $\hat P_{\munderbar U}$ in \eqref{2_isit2019}, i.e., $E_1(P_{\munderbar U},W)\geq E(P_{\munderbar U},W)$ where $ E(P_{\munderbar U},W)$ is given by
\begin{align}
E(P_{\munderbar U},W)
=
 \min_{i_1,i_2=1,2} \min_{\tau\in\left\{
 \{1\},\{2\},\{1,2\}
 \right\}} \max_{\rho\in[0,1]}
\min_{\hat{P}_{\munderbar X Y}\in\mathcal{P}_{\mathcal{\munderbar X}\times\mathcal{Y}}} 
D(\hat P_{\munderbar X Y}|| \Qu[1]{i_1}\Qu[2]{i_2}W)
\nonumber\hspace{3.5 em}\\
+ \rho D(\hat{P}_{\munderbar X Y}||\Qu[\tau]{i_\tau} \hat P_{X_{\tau^c}Y})
+\min_{\substack{\hat P_{\munderbar U}\in \mathcal P_{\munderbar{\mathcal U}}:\hat P_{\munderbar U}\in\Bu[1]{i_1},\hat P_{\munderbar U}\in\Bu[2]{i_2}}}
D(\hat P_{\munderbar U}||P_{\munderbar U})
-\rho H(\hat P_{U_\tau|U_{\tau^c}}).
 \label{3_isit2019}
\end{align}
In \eqref{3_isit2019}, the inner minimization problems over 
$\hat{P}_{\munderbar X Y}\in\mathcal{P}_{\mathcal{\munderbar X}\times\mathcal{Y}}$ and $\hat{P}_{\munderbar U}\in\mathcal{P}_{\mathcal{\munderbar U}}$, respectively lead to 
 the channel and source exponent functions. 
 The  minimization over $\hat P_{\munderbar U}$ is discussed in Proposition \ref{dif_Esi_isit2019}, while
to find channel exponent function, we use Lemma \ref{app_E0_0}.
By setting $\hat P_{XY}=\hat P_{\munderbar XY}$ and $Q=\Qu[\tau]{i_\tau}$ in Lemma \ref{app_E0_0}, the 
minimization over $\hat P_{\munderbar XY}$ in \eqref{3_isit2019}, is solved as
\begin{align}
\min_{\hat{P}_{\munderbar X Y}\in\mathcal{P}_{\mathcal{\munderbar X}\times\mathcal{Y}}} 
D(\hat P_{\munderbar X Y}|| \Qu[1]{i_1}\Qu[2]{i_2}W)
+ \rho D(\hat{P}_{\munderbar X Y}||\Qu[\tau]{i_\tau} \hat P_{X_{\tau^c}Y})
=E_0(\rho,\Qu[\tau]{i_\tau},W\Qu[\tau^c]{i_{\tau}^c}), \label{4_isit2019}
\end{align}
where $E_0(\cdot)$ is given by \eqref{E0_tau_isit2019}.

 Now, putting back the results obtained in equations  \eqref{4_isit2019} and  \eqref{bef_6_isit2019} into the respective minimization problems over $\hat P_{\munderbar X Y}$ and $\hat P_{\munderbar U}$ of \eqref{3_isit2019}, and defining
 \begin{align}
 f_{i_1,i_2} (\gamma_1,\gamma_2)=\min_{\tau\in\left\{
 \{1\},\{2\},\{1,2\}
 \right\}}\max_{\rho\in[0,1]}
E_0(\rho,\Qu[\tau]{i_\tau},W\Qu[\tau^c]{i_{\tau}^c})
-\EsCR[\tau,i_1,i_2]{}(\rho,P_{\munderbar U},\munderbar \gamma),
\label{f_small_isit2019}
 \end{align}
an alternative expression for \eqref{3_isit2019} is derived as
 \begin{align}
E(P_{\munderbar U},W)
=
\max_{\gamma_1,\gamma_2\in[0,1]}  \min_{i_1,i_2=1,2}  f_{i_1,i_2} (\gamma_1,\gamma_2)
,\label{2018.10.019_15.13_isit2019}
\end{align}
where in \eqref{2018.10.019_15.13_isit2019}, 
we optimized the exponent over $\gamma_\nu$ for $\nu=1,2$. We recall that 
since two source-message classes namely  $\Bu[\nu]{1}$, $\Bu[\nu]{2}$
and two input distributions $\Qu[\nu]{1}, \Qu[\nu]{2}$  are considered for each user $\nu=1,2$, there are four possible assignments where in
 \eqref{2018.10.019_15.13_isit2019} the optimal assignment of input distributions is considered.
 
In Appendix \ref{proof_prop2_isit}, we show that  for $\nu=1,2$, the function    $\max_{\rho\in[0,1]}
E_0(\rho,\Qu[\tau]{i_\tau},W\Qu[\tau^c]{i_{\tau}^c})
-\EsCR[\tau,i_1,i_2]{}(\rho,P_{\munderbar U},\munderbar \gamma)$  is non-decreasing 
with respect to $\gamma_\nu$  when $i_\nu=1$  and is  non-increasing  with respect to $\gamma_\nu$  when $i_\nu=2$.  Considering this fact,  to find the optimal $\munderbar \gamma$  maximizing \eqref{2018.10.019_15.13_isit2019}, we can use the same approach proposed in \cite[Proposition 2]{itw2018}. In other words,
the optimal $\gamma_1$ and $\gamma_2$ are the points where the minimum of all non-decreasing functions with respect to $\gamma_\nu$ 
are equal with the minimum of all non-increasing functions.

 \begin{proposition}
  \label{isit_prop1_20.10.2019}
 The optimal $\gamma_1^\star$ and $\gamma_2^\star$ maximizing  \eqref{2018.10.019_15.13_isit2019}  satisfy
 \begin{align}
 \left\{
 \begin{array}{rl}
\displaystyle\min_{i_{2}=1,2}
 f_{1,i_{2}}( \gamma_1^\star,\gamma_2^\star)=
\min_{i_{2}=1,2}
 f_{2,i_{2}}( \gamma_1^\star,\gamma_2^\star),
\\
\displaystyle \min_{i_{1}=1,2}
 f_{i_{1},1}( \gamma_1^\star,\gamma_2^\star)= \min_{i_{1}=1,2}
 f_{i_{1},2}( \gamma_1^\star,\gamma_2^\star) .\label{optimal_gamma_isit2019}
  \end{array}\right.
 \end{align}
If \eqref{optimal_gamma_isit2019} has no solutions, $\gamma_\nu^\star\in\{0,1\}$: if $ f_{1,i_{2}}(0, \gamma_2) > f_{2,i_{2}}(0, \gamma_2)$ then 
 $\gamma_1^\star=0$,  otherwise $\gamma_1^\star=1$; and 
 if $ f_{i_{1},1}(\gamma_1,0) > f_{i_{1},2}( \gamma_1,0)$, we have $\gamma_2^\star=0$, otherwise $\gamma_2^\star=1$.
\end{proposition}
\begin{IEEEproof}
See Appendix \ref{proof_prop2_isit}.
\end{IEEEproof}

Next, we show that the achievable exponent given by \eqref{2018.10.019_15.13_isit2019}, is greater than i.i.d random-coding exponent. 
\begin{proposition}
\label{LB_ISIT2019_bem}
The achievable exponent given by \eqref{2018.10.019_15.13_isit2019}
is greater than that achieved using only one input distribution for each user, i.e.,
\begin{align}
E(P_{\munderbar U},W)\geq \max_{i_1\in\{1,2\}} \max_{i_2\in\{1,2\}}
 \min_{\tau} 	F^{\rm L}_{\tau,i_\tau,i_{\tau^c}},
\label{lb_isit2019}
\end{align}
where
\begin{equation}
	F^{\rm L}_{\tau,i_\tau,i_{\tau^c}} = \max_{\rho\in[0,1]}
E_0(\rho,\Qu[\tau]{i_\tau},W\Qu[\tau^c]{i_{\tau^c}})
-E_{s,\tau}(\rho,P_{\munderbar  U}).
\label{eq:Fl_isit2019}
\end{equation}
\end{proposition}
Like \cite[Eq. (25)]{itw2018}, the lower bound in \eqref{lb_isit2019} selects the best 
i.i.d. random-coding exponent 
among the all four combinations of input distributions through $i_1$ and $i_2$.
\begin{IEEEproof}
See Appendix \ref{proof_LB_ISIT2019_bem}.
\end{IEEEproof}

 \section{Numerical Example}
 In this section, we present an example showing that
 using two input distributions for each user attains larger achievable exponent than the case where each user uses one input distribution.
 We consider
 two correlated discrete memoryless sources with alphabet $\mathcal U_\nu=\{1,2\}$ for $\nu=1,2$ where 
\begin{align}
    P_{\munderbar U}=\left (
     \begin{array}{cl}
0.0005\hspace{2em} 0.0095\\
0.0005\hspace{2em} 0.9895
     \end{array}\right).
\end{align}

  We also consider
 a discrete memoryless  MAC, similar to the one given by \cite[Eq. (31)]{itw2018}, with $\mathcal{X}_1=\mathcal{X}_2=\{1,\ldots,6\}$ and $|\mathcal{Y}|=4$.
Let $W$ be the transition probability of this channel, 
  \begin{align}
    W=\left(W_1^T, W_2^T, W_3^T, W_4^T, W_5^T, W_6^T\right)^T,
      \label{W_itw_final}
 \end{align}
 where 
 \begin{align}
W_{1}=\left( \begin{matrix} 
1-3k_1& k_1&k_1&k_1\cr
k_1&1-3k_1& k_1&k_1\cr
k_1&k_1&1-3k_1& k_1\cr
k_1&k_1& k_1&1-3k_1\cr
0.5-k_2& 0.5-k_2& k_2& k_2\cr
 k_2& k_2&0.5-k_2& 0.5-k_2
\end{matrix} \right),
 \end{align}
 for $k_1=0.045$ and $k_2=0.01$. $W_2$ and $W_3$ are $6\times4$ matrices whose rows are all the copy of $5^{\text{th}}$ and $6^{\text{th}}$ row of matrix $W_1$, respectively. $W_4$ is a $6\times4$ matrix with rows numbers 2, 3, 4, 1, 6, and 5 of $W_1$. Similarly, $W_5$ is a $6\times4$ matrix with rows numbers 3, 4, 1, 2, 5, and 6 of $W_1$ and $W_6$ is a $6\times4$ matrices with rows numbers 4, 1, 2, 3, 6, and 5 of $W_1$.
 
%

We observe that $W$ is a $36\times 4$ matrix where the transition probability  $W(y|x_1,x_2)$ is located at   row  $x_1+6(x_2-1)$ of matrix $W$, for
 $(x_1,x_2)\in\{1,2,...,6\}\times\{1,2,...,6\}$.
Recalling that each source has two classes and that four input distributions generate codewords, there are four possible assignments of input distributions to classes. 
 Among all possible permutations, we select the one that gives the highest exponent. Here, 
for user $\nu=1,2$, we consider the set of input distributions  $\big\{[0\hspace{.4em}0 \hspace{.4em}0\hspace{.4em}0\hspace{.4em}0.5
 \hspace{.4em}0.5],$ $[0.25\hspace{.4em}0.25 \hspace{.4em}0.25\hspace{.4em} 0.25 \hspace{.4em}0\hspace{.4em}0]\big\}$. 
For the channel given in \eqref{W_itw_final}, 
 the optimal assignment is
 \begin{align}
 \Qu[\nu]{1}=[0\hspace{.4em}0 \hspace{.4em}0\hspace{.4em}0\hspace{.4em}0.5
 \hspace{.4em}0.5],\hspace{3.5em} \label{eq:q1}\\
 \Qu[\nu]{2}=[0.25\hspace{.4em}0.25 \hspace{.4em}0.25\hspace{.4em} 0.25 \hspace{.4em}0\hspace{.4em}0],\label{eq:q2}
 \end{align}
for both $\nu=1,2$. 

\begin{figure}[!t]
\centering
\footnotesize
\input{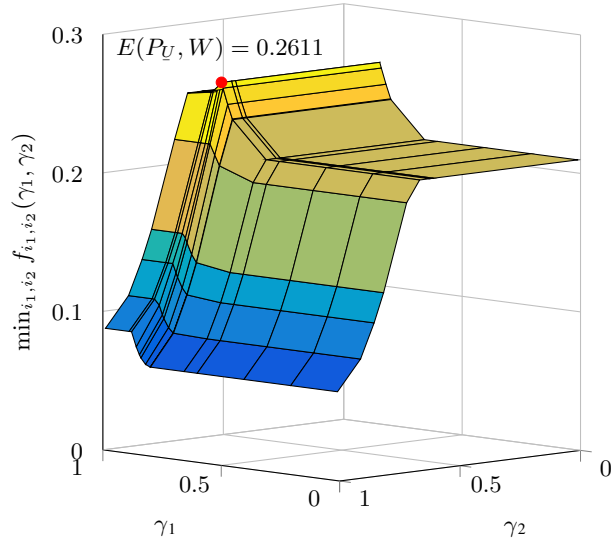}
\caption{$\min_{i_1,i_2}f_{i_1,i_2}( \gamma_1,\gamma_2)$ with respect to $\gamma_1$ and $\gamma_2$.}
\squeezeup
\label{fig_gamma}
\end{figure}

 For this example, from \eqref{optimal_gamma_isit2019},
we numerically  compute the optimal $\gamma_1^\star$ and $\gamma_2^\star$  maximizing \eqref{2018.10.019_15.13_isit2019} leading to $\gamma_1^\star=0.8469$ and $\gamma_2^\star=0.6581$. 
The message-dependent  exponent is derived as $E(P_{\munderbar U},W)=0.2611$, while  i.i.d. exponent for  the best assignment is derived as $0.2503$. Fig. \ref{fig_gamma} shows $\min_{i_1,i_2}f_{i_1,i_2}(\munderbar \gamma)$ with respect to $\gamma_1$ and $\gamma_2$. It can be seen that the maximum of  $\min_{i_1,i_2}f_{i_1,i_2}(\munderbar \gamma)$  is derived  at 
 $(0.8469,0.6581)$; however, the lower bound is obtained at $(1,0)$.

 \begin {table}[!t]
\begin{center}
\caption{ Values of $\max_{\rho\in [0,1]} E_0(\rho,\Qu[\tau]{i_\tau},W\Qu[\tau^c]{i_{\tau}^c})-
\EsCR[\tau,i_1,i_2]{}(\rho,P_{\underline U},\underline \gamma)$  with optimal thresholds $\gamma_1^\star=0.8469$  $\gamma_2^\star=0.6581$, for types of error $\tau$, and user classes $i_\tau$, $i_{\tau^c}$.}
\label{tab1_itw}            
\begin{tabular}{*{5}{c}}
\toprule
\multicolumn{5}{c}{$\hspace{6em}(i_1,i_2)$}
\\
\cline{2-5}
&(1,1)&(1,2)&(2,1)&(2,2)
\\
\midrule
 $\tau=\{1\}$& 0.3172  & 0.2735  &  0.3120  &  \cellcolor[gray]{0.9}0.2611
\\
 $\tau=\{2\}$&0.3986   & 0.4372 &  \cellcolor[gray]{0.9} 0.2611 &   0.4119
\\
 $\tau=\{1,2\}$& \cellcolor[gray]{0.9} 0.2611   & 0.2972  &  0.2630  &  0.2883
\\
\bottomrule
\end{tabular}
\end{center}
\end {table}
\begin {table}[!t]
\begin{center}  
\caption{Values of $F^{\rm L}_{\tau,i_\tau,i_{\tau^c}}$ in \eqref{eq:Fl_isit2019} for types of error $\tau$, and input distribution $\Qu[1]{i_1},\Qu[2]{i_2}$.}
\label{tab2_itw}    
\begin{tabular}{*{5}{c}} 
\toprule
&\Qu[1]{1},\Qu[2]{1}&\Qu[1]{1},\Qu[2]{2}&\Qu[1]{2},\Qu[2]{1}&\Qu[1]{2},\Qu[2]{2}
\\
\midrule
 $\tau=\{1\}$&0.2682  & \cellcolor[gray]{0.9} 0.0642  &  0.3120   & \cellcolor[gray]{0.9} 0.0879
\\
 $\tau=\{2\}$& 0.3986  &  0.3986  & \cellcolor[gray]{0.9} 0.2503   & 0.3696
\\
 $\tau=\{1,2\}$& \cellcolor[gray]{0.9} 0.2097  &  0.2097 &   0.2630   & 0.2360
 \\
\bottomrule
\end{tabular}
\end{center}
\end {table}

\appendices
\section{Proof of Lemma \ref{mac_partition_isit2019}}
\label{Proof_mac_partition}
Recalling $ \mathcal P_{\munderbar{\mathcal U}}^n$ is the set of  all empirical distributions on a joint vector in $\munderbar{\mathcal U}^n$, and $\mathcal T^n(\tilde P_{ \munderbar{U}})$ is the set of all joint sequences in $\munderbar{\mathcal U}^n$ with empirical distribution 
$\tilde P_{\munderbar U}$, 
$\munderbar{\mathcal U}^n$ can be  partitioned by all possible empirical distributions, i.e.,   
 $\munderbar{\mathcal U}^n=\bigcup_{\tilde P_{\munderbar U}\in \mathcal P_{\munderbar{\mathcal U}}^n} \mathcal T^n(\tilde P_{ \munderbar{U}})$. 
Since all $\munderbar{\bm u}$ belonging to the set $\mathcal T^n(\tilde P_{ \munderbar{U}})$ has the same probability, 
 the set  $\munderbar{\mathcal U}^n$ can be partitioned into two classes  $\Au[\nu]{1}$ and $\Au[\nu]{2}$ as
\begin{align}
\Au[\nu]{1}(\gamma_\nu)\triangleq\big\{\munderbar{\bm u}\in 
\bigcup_{\tilde P_{\munderbar U}\in \mathcal P_{\munderbar{\mathcal U}}^n} \mathcal T^n(\tilde P_{ \munderbar{U}})
 \,:\, P_{\bm U_\nu}^n(\bm u_\nu)\geq \gamma_\nu^n\big\}\label{A1_isit2019_nef},\\
\Au[\nu]{2}(\gamma_\nu)\triangleq\big\{\munderbar{\bm u}\in \bigcup_{\tilde P_{\munderbar U}\in \mathcal P_{\munderbar{\mathcal U}}^n}\mathcal T^n(\tilde P_{ \munderbar{U}})\,:\, P_{\bm U_\nu}^n(\bm u_\nu)< \gamma_\nu^n\big\},
\label{A2_isit2019_nef}
\end{align}
for a given $\gamma_\nu\in[0,1]$ where $\nu=1,2$.

By letting $\nu^c$ as the complement of  $\nu\in\{1,2\}$,  and noting that for 
 $\munderbar{\bm u}\in \mathcal T^n(\tilde P_{\munderbar U})$, the sequence $\bm u_\nu$
contains exactly $n\sum_{u_{\nu^c}}\tilde P_{\munderbar U} (\munderbar u)$
occurrences of $u_\nu$, the
probability of  $\bm u_\nu$ is $P_{\bm U_\nu} (\bm u_\nu)=\prod_{u_\nu\in \mathcal U_\nu}P_{ U_\nu}( u_\nu)^{n \sum_{u_{ \nu^c}} \tilde P_{\munderbar U }(\munderbar u)}$.
Let $\mathcal T^n(\tilde  P_{\munderbar U}) \subseteq \Au[\nu]{1}$, for $\munderbar{\bm u}^n\in\mathcal T^n(\tilde P_{\munderbar U})$, the condition $P_{\bm U_\nu}^n(\bm u_\nu)\geq \gamma_\nu^n$ can be written as $\prod_{u_\nu\in \mathcal U_\nu}P_{ U_\nu}( u_\nu)^{n \sum_{u_{\bar \nu}} \tilde P_{\munderbar U }(\munderbar u)}\geq \gamma_\nu^n$ where by taking logarithm from both sides, it is simplified as
$\sum_{\munderbar u}\tilde P_{\munderbar U} (\munderbar u) \log P_{U_\nu}(u_\nu)\geq \log(\gamma_\nu)$.
Using the same reasoning for $\mathcal T^n(\tilde  P_{\munderbar U}) \subseteq \Au[\nu]{2}$, the sets $\Au[\nu]{1}$ and $\Au[\nu]{2}$ can be rewritten as 
\begin{align}
\Au[\nu]{1}(\gamma_\nu)=\left\{ 
\tilde P_{\munderbar U}\in \mathcal P_{\munderbar{\mathcal U}}^n
:\sum_{\munderbar u}\tilde P_{\munderbar U} (\munderbar u) \log P_{ U_\nu}(u_\nu)\geq \log(\gamma_\nu)
\right\}\label{A1_itw2018_primal_form},\\
\Au[\nu]{2}(\gamma_\nu)=\left\{ 
\tilde P_{\munderbar U}\in \mathcal P_{\munderbar{\mathcal U}}^n
:\sum_{\munderbar u}\tilde P_{\munderbar U} (\munderbar u) \log P_{ U_\nu}(u_\nu)< \log(\gamma_\nu)
\right\},
\label{A2_itw2018_primal_form}
\end{align}
where in \eqref{A1_itw2018_primal_form} and 
\eqref{A2_itw2018_primal_form}, we express $\Au[\nu]{1}$ and $\Au[\nu]{2}$ in terms of  empirical distributions.

As $n$ tends to infinity, since the set of all empirical distributions is dense
in the set of all possible probability distributions $\mathcal P_{\munderbar{\mathcal U}}$, 
the sets $\Au[\nu]{1}$ and $\Au[\nu]{2}$, respectively tend to $\Bu[\nu]{1}$ and $\Bu[\nu]{2}$ given by \eqref{B1_isit2019} and \eqref{B2_isit2019}, and hence
Lemma \ref{mac_partition_isit2019} is proved.

\section{Proof of Proposition \ref{dif_Esi_isit2019}}
\label{Proof_dif_Esi_isit2019}
To prove Proposition \ref{dif_Esi_isit2019}, we start by finding the dual form of the following problem:
\begin{align}
\min_{\substack{\hat P_{\munderbar U}\in \mathcal P_{\munderbar{\mathcal U}}:\hat P_{\munderbar U}\in\Bu[1]{i_1},\hat P_{\munderbar U}\in\Bu[2]{i_2}}}D(\hat P_{\munderbar U}||P_{\munderbar U})
-\rho H(\hat P_{U_\tau|U_{\tau^c}}),
\label{first_pr_isit2019_n1}
\end{align}
by applying Lagrange duality to 
the minimization problem.
We use $\Larezou[1]{}$ and $\Larezou[2]{}$ as the Lagrange multipliers, respectively
associated with the constraints $\hat P_{\munderbar U}\in\Bu[1]{i_1}$ and  $\hat P_{\munderbar U}\in\Bu[2]{i_2}$.

Firstly, we simplify
the objective function in \eqref{first_pr_isit2019_n1}.
Since
 $D(\hat P_{U_{\tau^c}}||V_{U_{\tau^c}})\geq 0$, for any $V_{U_{\tau^c}}\in \mathcal P_{U_{\tau^c}}$, we have 
\begin{equation}\sum_{\munderbar u}\hat P_{\munderbar U} (\munderbar u)\log \hat P_{U_{\tau^c}}(u_{\tau^c})
\geq 
\sum_{\munderbar u}\hat P_{\munderbar U} (\munderbar u)\log 
V_{U_{\tau^c}} (u_{\tau^c})
.\end{equation}
Multiplying  both sides of the inequality 
by $-1$
and adding $-H(\hat P_{\munderbar U})$ to both sides, we find that
\begin{equation}\sum_{\munderbar u}\hat P_{\munderbar U} (\munderbar u)\log \frac{\hat P_{\munderbar U} (\munderbar u)}{\hat P_{U_{\tau^c}} (u_{\tau^c})}
\leq \sum_{\munderbar u}\hat P_{\munderbar U} (\munderbar u)\log \frac{\hat P_{\munderbar U} (\munderbar u)}{V_{U_{\tau^c}} (u_{\tau^c})}
.\end{equation}
Recalling the definition of $H(\hat P_{U_\tau|U_{\tau^c}})$, the left hand side of the inequality is $-H(\hat P_{U_\tau|U_{\tau^c}})$, and therefore
$-H(\hat P_{U_\tau|U_{\tau^c}})
\leq \sum_{\munderbar u}\hat P_{\munderbar U} (\munderbar u)\log \frac{\hat P_{\munderbar U} (\munderbar u)}{V_{U_{\tau^c}} (u_{\tau^c})}
$. From this inequality, we conclude that the right hand side of the inequality is always greater than $-H(\hat P_{U_\tau|U_{\tau^c}})$ and only is equal to   $-H(\hat P_{U_\tau|U_{\tau^c}})$  when $V_{U_{\tau^c}} (u_{\tau^c})=P_{U_{\tau^c}} (u_{\tau^c})$ for all values of $u_{\tau^c}\in \mathcal U_{\tau^c}$, i.e.,
$\min_{V_{U_{\tau^c}}}  \sum_{\munderbar u}\hat P_{\munderbar U} (\munderbar u)\log \frac{\hat P_{\munderbar U} (\munderbar u)}{V_{U_{\tau^c}} (u_{\tau^c})}=- H(\hat P_{U_\tau|U_{\tau^c}})$. By applying this fact to the objective function in \eqref{first_pr_isit2019_n1}, we obtain
\begin{align}
D(\hat P_{\munderbar U}||P_{\munderbar U})
-\rho H(\hat P_{U_\tau|U_{\tau^c}})=\min_{V_{U_{\tau^c}}\in \mathcal P_{U_{\tau^c}}} 
D(\hat P_{\munderbar U}||P_{\munderbar U})
+\rho \sum_{\munderbar u}\hat P_{\munderbar U} (\munderbar u)\log \frac{\hat P_{\munderbar U} (\munderbar u)}{V_{U_{\tau^c}} (u_{\tau^c})}. \label{10.08.2018_16.47}
\end{align}
 Applying  \eqref{10.08.2018_16.47} to  the objective function of 
 \eqref{first_pr_isit2019_n1}, we find that  
\begin{align}
\min_{\substack{\hat P_{\munderbar U}\in \mathcal P_{\munderbar{\mathcal U}}:\hat P_{\munderbar U}\in\Bu[1]{i_1},\hat P_{\munderbar U}\in \Bu[2]{i_2}
}}D(\hat P_{\munderbar U}||P_{\munderbar U})
-\rho H(\hat P_{U_\tau|U_{\tau^c}})
\hspace{14em}\nonumber\\
=
\min_{V_{U_{\tau^c}}\in \mathcal P_{U_{\tau^c}}}\min_{\substack{\hat P_{\munderbar U}\in \mathcal P_{\munderbar{\mathcal U}}:\hat P_{\munderbar U}\in\Bu[1]{i_1},\hat P_{\munderbar U}\in \Bu[2]{i_2}
}}
D(\hat P_{\munderbar U}||P_{\munderbar U})
+\rho \sum_{\munderbar u}\hat P_{\munderbar U} (\munderbar u)\log \frac{\hat P_{\munderbar U} (\munderbar u)}{V_{U_{\tau^c}} (u_{\tau^c})}.
 \label{10.08.2018_17.12_n1}
\end{align}

Next, 
we apply  Lagrange duality theory to the inner minimization over $\hat{P}_{\munderbar U}$ in
 \eqref{10.08.2018_17.12_n1}.
 Considering  the constraints $\hat P_{\munderbar U}\in\Bu[1]{i_1}$ and $\hat P_{\munderbar U}\in\Bu[2]{i_2}$ and in view of definitions \eqref{B1_isit2019} and \eqref{B2_isit2019},
the Lagrangian associated with the primal is
given by
\begin{align}
\Lambda(V_{U_{\tau^c}},\hat{P}_{\munderbar U },\theta,\Larezou[1]{},\Larezou[2]{})=D(\hat P_{\munderbar U }|| P_{\munderbar U})
+\rho \sum_{\munderbar u}\hat P_{\munderbar U} (\munderbar u)\log \frac{\hat P_{\munderbar U} (\munderbar u)}{V_{U_{\tau^c}} (u_{\tau^c})}
+\theta\left(1-\sum_{\munderbar u} \hat{P}_{\munderbar U }(\munderbar u)\right)
\hspace{2em}\nonumber\\
+(-1)^{i_1}\Larezou[1]{}\left(\sum_{\munderbar u,\munderbar x,y}\hat P_{\munderbar U}(\munderbar u)\log P_{U_1}(u_1)-\log\gamma_{1}\right)
+(-1)^{i_2}\Larezou[2]{}\left(\sum_{\munderbar u}\hat P_{\munderbar U}(\munderbar u)\log P_{U_2}(u_2)-\log\gamma_{2}\right)
,\label{20.09.2018_11.17_n1}
\end{align}
where $\Larezou[1]{}$, $\Larezou[2]{}$ and 
$\theta$ are respectively the Lagrange multipliers for the inequality constraints $\hat P_{\munderbar U}\in\Bu[1]{i_1}$, $\hat P_{\munderbar U}\in\Bu[2]{i_2}$ and the sum of any probability distribution over its alphabet is one.

Noting that
the objective function and the inequalities constrains in \eqref{B1_isit2019} and \eqref{B2_isit2019} are convex with respect to $\hat P_{\munderbar U }$, and the equality constraint is affine, 
strong duality conditions are satisfied. Thus, 
the primal optimal objective and the dual optimal objective are equal,
\begin{align}
\min_{V_{U_{\tau^c}}\in \mathcal P_{U_{\tau^c}}}\min_{\substack{\hat P_{\munderbar U}\in \mathcal P_{\munderbar{\mathcal U}}:\hat P_{\munderbar U}\in\Bu[1]{i_1}, \hat P_{\munderbar U}\in \Bu[2]{i_2}
}}
D(\hat P_{\munderbar U}||P_{\munderbar U})
+\rho \sum_{\munderbar u}\hat P_{\munderbar U} (\munderbar u)\log \frac{\hat P_{\munderbar U} (\munderbar u)}{V_{U_{\tau^c}} (u_{\tau^c})}
\hspace{3em}\nonumber\\
=
\min_{V_{U_{\tau^c}}\in \mathcal P_{U_{\tau^c}}}
\max_{\Larezou[1]{}\geq 0,\Larezou[2]{}\geq 0}
\max_{\theta}
\min_{\hat P_{\munderbar U}}
\Lambda(V_{U_{\tau^c}},\hat{P}_{\munderbar U },\theta,\Larezou[1]{},\Larezou[2]{})\label{10.09.2018_13.20_n1},
\end{align}
where we recall that for $\nu=1,2$,
the condition $\Larezou[\nu]{}\geq 0$ 
 in \eqref{10.09.2018_13.20_n1} is
 associated
with inequality constraint $(-1)^{i_\nu}\left(
\sum_{\munderbar u}\hat P_{\munderbar U} ( \munderbar u)\log P_{U_\nu}(u_\nu)-\log(\gamma_\nu)< 0\right)$.

Since  strong duality holds, in view of 
the KKT conditions, by setting  $\frac{\partial \Lambda}{\partial \hat{P}_{\munderbar U }}=0$, and applying the constraint 
$\sum_{\munderbar u}\hat P_{\munderbar U }(\munderbar u)=1$, 
we obtain
\begin{align}
\Lambda(V_{U_{\tau^c}},\Larezou[1]{},\Larezou[2]{})=
-(-1)^{i_1 }\Larezou[1]{}\log\gamma_{1}-(-1)^{i_2 }
\Larezou[2]{}
\log\gamma_{2}
\hspace{12em}\nonumber\\
-(1+\rho)\log\left(
\sum_{\munderbar u} P_{\munderbar U}(\munderbar u)^{\frac{1}{1+\rho}}
P_{U_1}(u_1)^{-\frac{(-1)^{i_1}\Larezou[1]{}}{1+\rho}}
P_{U_2}(u_2)^{-\frac{(-1)^{i_2}\Larezou[2]{}}{1+\rho}}
V_{U_{\tau^c}}(u_{\tau^c})^{\frac{\rho}{1+\rho}}
\right),
\label{2.10.09.2018_13.12_n1}
\end{align}
where $\Lambda(V_{U_{\tau^c}},\Larezou[1]{}, \Larezou[2]{})=\max_{\theta}\min_{\hat P_{\munderbar U}}\Lambda(V_{U_{\tau^c}},\hat P_{\munderbar U},\theta,\Larezou[1]{},\Larezou[2]{})$. Inserting
$\Lambda(V_{U_{\tau^c}},\Larezou[1]{}, \Larezou[2]{})$ derived in \eqref{2.10.09.2018_13.12_n1}
into  \eqref{10.09.2018_13.20_n1}, we find
\begin{align}
\min_{V_{U_{\tau^c}}}\min_{\substack{\hat P_{\munderbar U}\in \mathcal P_{\munderbar{\mathcal U}}:\hat P_{\munderbar U}\in\Bu[1]{i_1}, \hat P_{\munderbar U}\in\Bu[2]{i_2}
}}
D(\hat P_{\munderbar U}||P_{\munderbar U})
+\rho \sum_{\munderbar u}\hat P_{\munderbar U} (\munderbar u)\log \frac{\hat P_{\munderbar U} (\munderbar u)}{V_{U_{\tau^c}} (u_{\tau^c})}
\hspace{5em}\nonumber\\
=
\min_{V_{U_{\tau^c}}\in \mathcal P_{U_{\tau^c}}}
\max_{\Larezou[1]{}\geq 0,\Larezou[2]{}\geq 0}
\Lambda(V_{U_{\tau^c}},\Larezou[1]{},\Larezou[2]{})\label{10.09.2018_13.22_n1}.
\end{align}

We note that in \eqref{10.09.2018_13.22_n1},
$V_{U_{\tau^c}}\in \mathcal P_{U_{\tau^c}}$ and 
$\Larezou[\nu]{}\in [0,+\infty)$ for $\nu=1,2$.
Since
 $\mathcal P_{U_{\tau^c}}$ is  a compact convex set, $[0,+\infty)$ is 
a convex  set, $\Lambda(V_{U_{\tau^c}},\Larezou[1]{},\Larezou[2]{})$ is a concave function over $\Larezou[1]{}$ and $\Larezou[2]{}$ on
$[0,+\infty)$ and  convex over 
$V_{U_{\tau^c}}$
on  $\mathcal P_{U_{\tau^c}}$, \eqref{10.09.2018_13.22_n1} satisfies
the 
Sion's minimax theorem \cite{sion}. Thus, we may swap 
the maximization over $\Larezou[\nu]{}$ with minimization over $V_{U_{\tau^c}}$ which leads to
\begin{align}
\min_{V_{U_{\tau^c}}}\min_{\substack{\hat P_{\munderbar U}\in \mathcal P_{\munderbar{\mathcal U}}:\\\hat P_{\munderbar U}\in\Bu[1]{i_1}, \hat P_{\munderbar U}\in\Bu[2]{i_2}
}}
D(\hat P_{\munderbar U}||P_{\munderbar U})
+\rho \sum_{\munderbar u}\hat P_{\munderbar U} (\munderbar u)\log \frac{\hat P_{\munderbar U} (\munderbar u)}{V_{U_{\tau^c}} (u_{\tau^c})}
\hspace{8em}\nonumber\\
=
\max_{\Larezou[1]{}\geq 0,\Larezou[2]{}\geq 0}
\min_{V_{U_{\tau^c}}\in \mathcal P_{U_{\tau^c}}}
\Lambda(V_{U_{\tau^c}},\Larezou[1]{},\Larezou[2]{})\label{10.09.2018_13.22_n1_zh}.
\end{align}

Next, to solve the minimization over $V_{U_{\tau^c}}$ in the right hand side of \eqref{10.09.2018_13.22_n1_zh}, 
by inserting $\Lambda(\cdot)$ given by
\eqref{2.10.09.2018_13.12_n1} into 
\eqref{10.09.2018_13.22_n1_zh}, we find that
\begin{align}
\min_{V_{U_{\tau^c}}}\min_{\substack{\hat P_{\munderbar U}\in \mathcal P_{\munderbar{\mathcal U}}:\\\hat P_{\munderbar U}\in\Bu[1]{i_1}, \hat P_{\munderbar U}\in\Bu[2]{i_2}
}}
D(\hat P_{\munderbar U}||P_{\munderbar U})
+\rho \sum_{\munderbar u}\hat P_{\munderbar U} (\munderbar u)\log \frac{\hat P_{\munderbar U} (\munderbar u)}{V_{U_{\tau^c}} (u_{\tau^c})}
=\max_{\Larezou[1]{}\geq 0,\Larezou[2]{}\geq 0}
\hspace{8em}\nonumber\\
-(1+\rho)\log\left(\max_{V_{U_{\tau^c}}\in \mathcal P_{U_{\tau^c}}}
\sum_{\munderbar u} P_{\munderbar U}(\munderbar u)^{\frac{1}{1+\rho}}
\left(\frac{P_{U_1}(u_1)}{\gamma_1}\right)^{-\frac{(-1)^{i_1}\Larezou[1]{}}{1+\rho}}
\left(\frac{P_{U_2}(u_2)}{\gamma_2}\right)^{-\frac{(-1)^{i_2}\Larezou[2]{}}{1+\rho}}
V_{U_{\tau^c}}(u_{\tau^c})^{\frac{\rho}{1+\rho}}
\right),\label{isit2019_tamom_sho}
\end{align}
where in \eqref{isit2019_tamom_sho},   we moved the minimization over $V_{U_{\tau^c}}$ inside the logarithm
as the logarithm is an
increasing function.
In addition, we used the fact that 
$-\log(\gamma)-(1+\rho)\log(a)= -(1+\rho)\log(\frac{a}{\gamma^{-\frac{1}{1+\rho}}})$.

 Now, 
 we can apply 
 Lemma \ref{Vyyy} of Appendix \ref{ham_kare o hich kare} into the optimization problem given by the right hand side of \eqref{isit2019_tamom_sho}.
 By defining \begin{equation} e(u_{\tau^c})=\sum_{u_\tau}  P_{\munderbar U}(\munderbar u)^{\frac{1}{1+\rho}}
\left(\frac{P_{U_1}(u_1)}{\gamma_1}\right)^{-\frac{(-1)^{i_1}\Larezou[1]{}}{1+\rho}}
\left(\frac{P_{U_2}(u_2)}{\gamma_2}\right)^{-\frac{(-1)^{i_2}\Larezou[2]{}}{1+\rho}}, \end{equation}
we let $V_Y(y)=V_{U_{\tau^c}}(u_{\tau^c})$ and $e(y)=e(u_{\tau^c})$ in Lemma \ref{Vyyy}. Thus,  the optimal 
$V_{U_{\tau^c}}$ is derived as
\begin{align}
V_{U_{\tau^c}}(u_{\tau^c})=\frac{\displaystyle\left(\sum_{u_\tau} P_{\munderbar U}(\munderbar u)^{\frac{1}{1+\rho}}
P_{U_1}(u_1)^{-\frac{(-1)^{i_1}\Larezou[1]{}}{1+\rho}}
P_{U_2}(u_2)^{-\frac{(-1)^{i_2}\Larezou[2]{}}{1+\rho}}\right)^{1+\rho}}
{\displaystyle\sum_{\bar u_{\tau^c}}
\left(\sum_{\bar u_\tau} P_{\munderbar U}(\bar u_1,\bar u_2)^{\frac{1}{1+\rho}}
P_{U_1}(\bar u_1)^{-\frac{(-1)^{i_1}\Larezou[1]{}}{1+\rho}}
P_{U_2}(\bar u_2)^{-\frac{(-1)^{i_2}\Larezou[2]{}}{1+\rho}}\right)^{1+\rho}
}.
\label{26.10.2019_17.15_n1}
\end{align}
In addition, in view of \eqref{Vy_13march} in
Lemma \ref{Vyyy},
the optimization  problem over $V_{U_{\tau^c}}$ in \eqref{isit2019_tamom_sho}
is equal to
 $\left( \sum_{u_{\tau^c}} e(u_{\tau^c})^{1+\rho}\right)^{\frac{1}{1+\rho}}$, i.e.,
\begin{align}
\min_{V_{U_{\tau^c}}}\min_{\substack{\hat P_{\munderbar U}\in \mathcal P_{\munderbar{\mathcal U}}:\hat P_{\munderbar U}\in\Bu[1]{i_1}, \hat P_{\munderbar U}\in\Bu[2]{i_2}
}}
D(\hat P_{\munderbar U}||P_{\munderbar U})
+\rho \sum_{\munderbar u}\hat P_{\munderbar U} (\munderbar u)\log \frac{\hat P_{\munderbar U} (\munderbar u)}{V_{U_{\tau^c}} (u_{\tau^c})}
=
\hspace{6em}\nonumber\\
\max_{\Larezou[1]{}\geq 0,\Larezou[2]{}\geq 0}-\log\left(
\sum_{u_{\tau^c}} \left(
\sum_{u_\tau}  P_{\munderbar U}(\munderbar u)^{\frac{1}{1+\rho}}
\left(\frac{P_{U_1}(u_1)}{\gamma_1}\right)^{-\frac{(-1)^{i_1}\Larezou[1]{}}{1+\rho}}
\left(\frac{P_{U_2}(u_2)}{\gamma_2}\right)^{-\frac{(-1)^{i_2}\Larezou[2]{}}{1+\rho}}
\right)^{1+\rho}
\right),
\label{10.09.2018_16.07_n1}
\end{align}
 where considering \eqref{10.08.2018_17.12_n1}, by replacing the left hand side of  \eqref{10.08.2018_17.12_n1}
 with the left hand side of 
 \eqref{10.09.2018_16.07_n1}, 
 we conclude
  \begin{align}
\min_{\substack{\hat P_{\munderbar U}\in \mathcal P_{\munderbar{\mathcal U}}:\hat P_{\munderbar U}\in\Bu[1]{i_1},\hat P_{\munderbar U}\in\Bu[2]{i_2}}}D(\hat P_{\munderbar U}||P_{\munderbar U})
-\rho H(\hat P_{U_\tau|U_{\tau^c}})=
\hspace{13em}\nonumber\\
-\min_{\Larezou[1]{}\geq 0,\Larezou[2]{}\geq 0}\log\left(
\sum_{u_{\tau^c}} \left(
\sum_{u_\tau}  P_{\munderbar U}(\munderbar u)^{\frac{1}{1+\rho}}
\left(\frac{P_{U_1}(u_1)}{\gamma_1}\right)^{-\frac{(-1)^{i_1}\Larezou[1]{}}{1+\rho}}
\left(\frac{P_{U_2}(u_2)}{\gamma_2}\right)^{-\frac{(-1)^{i_2}\Larezou[2]{}}{1+\rho}}
\right)^{1+\rho}
\right),
\label{cs1_isit20198}
\end{align}
 where in \eqref{cs1_isit20198}, we used the fact that $\max_\lambda- f(\lambda)=-\min_\lambda f(\lambda)$.
 Comparing \eqref{cs1_isit20198} with \eqref{6.1_isit2019}  concludes  the proof.

\section{Proof of Lemma \ref{remove_one_mini_isit2019}}
\label{proof_remove_one_mini_isit2019}
To bound  $E_1(P_{\munderbar U},W)$ in
\eqref{primal_exp_isit2019},
 we first
apply Lemma \ref{removing one minimization_noncondi} in Appendix \ref{ham_kare o hich kare} to \eqref{primal_exp_isit2019}.
 By setting $T=U_\tau$, $Z=U_{\tau^c}$,
  $U=\munderbar U$, $X=X_\tau$,  $W=Q_{\tau^c,\hat P_{U_{\tau^c}}}W$
 and $Y=X_{\tau^c}Y$ in 
Lemma \ref{removing one minimization_noncondi}, 
 an achievable exponent given by \eqref{primal_exp_isit2019} is bounded  as
\begin{align}
E_1(P_{\munderbar U},W)\geq
 \min_{\tau}
\min_{\hat{P}_{\munderbar U}
 \in\mathcal{P}_{ \mathcal{\munderbar U}}
 }\min_{\hat{P}_{\munderbar  X Y}
 \in\mathcal{P}_{ \mathcal{\munderbar X}\times\mathcal{Y}}
 } D(\hat P_{\munderbar U }|| P_{\munderbar U})
+D(\hat P_{\munderbar  X Y}||Q_{1,\hat P_{U_1}}Q_{2,\hat P_{U_2}}W)
\hspace{5em}\nonumber\\
+ \bigg [
D(\hat{P}_{\munderbar  X Y}|| Q_{\tau,\hat P_{U_\tau}}\hat{P}_{X_{\tau^c}Y})
-H(\hat P_{U_\tau|U_{\tau^c}})
\bigg]^+.
\label{2019.01.08_12.39_isit2019}
\end{align}
By using 
the identity  $\max\{0,a\}=\max_{\rho\in[0,1]} \rho a$, \eqref{2019.01.08_12.39_isit2019} is simplified as
\begin{align}
E_1(P_{\munderbar U},W)\geq \min_{\tau}
\min_{\hat{P}_{\munderbar X Y}
\in\mathcal{P}_{\mathcal{\munderbar X}\times\mathcal{Y}}
} \min_{\substack{\hat P_{\munderbar U}
\in \mathcal P_{\munderbar{\mathcal U}}
}}D(\hat P_{\munderbar U}||P_{\munderbar U})
+
D(\hat P_{\munderbar X Y}|| \Qu[1]{\hat P_{U_1}}\Qu[2]{\hat P_{U_2}}W)
\hspace{5em}\nonumber\\
+\max_{\rho\in[0,1]}
 \rho D(\hat{P}_{ \munderbar X Y}||\Qu[\tau]{\hat P_{U_\tau}}\hat P_{X_{\tau^c}Y})
-\rho H(\hat P_{U_\tau|U_{\tau^c}})
,\label{dovom_isit2019}
\end{align}
which concludes the proof.

\section{Proof of Proposition \ref{isit_prop1_20.10.2019}}
\label{proof_prop2_isit}
We start by proving the following Lemma.
\begin{lemma}
\label{dec-inc with gamm}
Let $E_0(\rho)$ be a continuous function of $\rho$.
Considering $\EsCR[\tau,i_1,i_2]{}(\cdot)$ given by \eqref{6.1_isit2019},
for $\nu=1,2$, the function
 $\max_{\rho}E_0(\rho)-\EsCR[\tau,i_1,i_2]{}(\cdot)\big|_{i_\nu=1}$ is non-decreasing with respect to $\gamma_\nu$, and the function
 $\max_{\rho}E_0(\rho)-\EsCR[\tau,i_1,i_2]{}(\cdot)\big|_{i_\nu=2}$ is non-increasing with respect to $\gamma_\nu$.
\end{lemma}
\begin{IEEEproof}
For $\nu=1,2$, from \eqref{B1_isit2019} and  \eqref{B2_isit2019}, we note that by letting $\gamma_\nu^\prime > \gamma_\nu^{\prime\prime}$, we have $\Cua[\nu]{1}(\gamma_\nu^\prime) \subseteq \Cua[\nu]{1}(\gamma_\nu^{\prime\prime})$ and $\Cua[\nu]{2}(\gamma_\nu^\prime) \supseteq \Cua[\nu]{2}(\gamma_\nu^{\prime\prime})$. 
Thus, for all $\rho\in[0,1]$ by letting $i_\nu=1$ in \eqref{bef_6_isit2019}, 
we conclude that 
for $\gamma_\nu^\prime$ the
 minimization problem of \eqref{bef_6_isit2019}  is done over smaller  set than for $\gamma_\nu^{\prime\prime}$,
 which leads to   $-\EsCR[\tau,i_1,i_2]{}(\cdot,\gamma_\nu^\prime)\big|_{i_\nu=1}\geq -\EsCR[\tau,i_1,i_2]{}(\cdot,\gamma_\nu^{\prime\prime})\big|_{i_\nu=1}$ for all values of $\rho\geq 0$. Similarly, for 
 $i_\nu=2$, since the minimization problem 
of \eqref{bef_6_isit2019} for $\gamma_\nu^\prime$ is done over larger  set than $\gamma_\nu^{\prime\prime}$, we have   
$-\EsCR[\tau,i_1,i_2]{}(\cdot,\gamma_\nu^\prime)\big|_{i_\nu=2}\leq -\EsCR[\tau,i_1,i_2]{}(\cdot,\gamma_\nu^{\prime\prime})\big|_{i_\nu=2}$.

 Hence, let
 $\nu^c$ be the complement index of $\nu\in\{1,2\}$ and 
 $E_0(\rho)$ be a function of $\rho$.  
 For given $\gamma_{\nu^c}$, we assume
 $\gamma_\nu^\prime > \gamma_\nu^{\prime\prime}$. Thus, regardless of the value of $i_{\nu^c}$,
 the maximum of $E_0(\rho)-\EsCR[\tau,i_1,i_2]{}(\cdot,\gamma_\nu^\prime)\big|_{i_\nu=1}$ is not smaller than 
 $E_0(\rho)-\EsCR[\tau,i_1,i_2]{}(\cdot,\gamma_\nu^{\prime\prime})\big|_{i_\nu=1}$, and therefore $\max_{\rho}E_0(\rho)-\EsCR[\tau,i_1,i_2]{}(\cdot)\big|_{i_\nu=1}$ is non-decreasing with respect to $\gamma_\nu$. The same reasoning allows us to conclude that
  $\max_{\rho}E_0(\rho)-\EsCR[\tau,i_1,i_2]{}(\cdot)\big|_{i_\nu=2}$ is non-increasing with respect to $\gamma_\nu$.
\end{IEEEproof}
Now, in view of \eqref{f_small_isit2019}, we define
 $F_{\tau,i_1,i_2}(\munderbar \gamma)$ as
\begin{align}
F_{\tau,i_1,i_2}(\munderbar \gamma)=\max_{\rho\in [0,1]} E_0(\rho,\Qu[\tau]{i_\tau},W\Qu[\tau^c]{i_{\tau}^c})-\EsCR[\tau,i_1,i_2]{}(\rho,P_{\munderbar U},\munderbar \gamma),
\end{align}
where $f_{i_1,i_2}(\munderbar \gamma)=\min_\tau
F_{\tau,i_1,i_2}(\munderbar \gamma)$.
We note that $F_{\tau,i_1,i_2}(\munderbar \gamma)$ 
is of the form $\max_{\rho\in [0,1]} E_0(\rho)-\EsCR[\tau,i_1,i_2]{}(\rho,P_{\munderbar U},\munderbar \gamma)$ in  Lemma \ref{dec-inc with gamm}.
In view of Lemma  \ref{dec-inc with gamm}, 
$F_{\tau,1,i_2}$ and  $F_{\tau,2,i_2}$ are respectively non-decreasing and non-increasing
with respect to $\gamma_1$. Similarly, regardless of the value of $i_1$, 
$F_{\tau,i_1,1}$ and  $F_{\tau,i_1,2}$ are respectively non-decreasing and non-increasing
with respect to $\gamma_2$. 

Considering $f_{i_1,i_2}(\munderbar \gamma)=\min_\tau
F_{\tau,i_1,i_2}(\munderbar \gamma)$, by
applying the fact that the  minimum of monotonic functions is monotonic, $f_{i_1,i_2}(\munderbar \gamma)$ in \eqref{f_small_isit2019} is non-decreasing (resp. non-increasing)
with respect to $\gamma_\nu$ when $i_\nu=1$ (resp. $i_\nu=2$) for $\nu=1,2$.

Next, to find the optimal $\munderbar \gamma$ maximizing \eqref{2018.10.019_15.13_isit2019} we express 
$E(P_{\munderbar U},W)$ as
\begin{align}
\max_{\gamma_1}\max_{\gamma_2}\min_{i_2} \min_{i_1}
f_{i_1,i_2}(\munderbar \gamma),
\end{align}
where for a fixed $\gamma_1$, the optimization problem  $\max_{\gamma_2}\min_{i_2} \min_{i_1}
f_{i_1,i_2}(\munderbar \gamma)$ 
satisfies \cite[Lemma 5]{arzou_arxiv}
with $\gamma=\gamma_2$, $i=i_2$, and $k_i(\gamma)=\min_{i_1} f_{i_1,i}(\gamma_1,\gamma)$.  Therefore, the optimal $\gamma_2^\star$ satisfies 
 \begin{align}
\min_{i_1=1,2} f_{i_1,1}(\gamma_1, \gamma_2^\star)=\min_{i_1=1,2} f_{i_1,2}(\gamma_1, \gamma_2^\star),\label{isit2019_copy1_itw}
\end{align}
 whenever \eqref{isit2019_copy1_itw} has solution. Otherwise, we have $\gamma_2^\star=0$ when $f_{i_1,1}(\gamma_1, 0)> f_{i_1,2}(\gamma_1,0)$, or $\gamma_2^\star=1$ when $f_{i_1,1}(\gamma_1, 0)\leq  f_{i_1,2}(\gamma_1,0)$.

Now, applying $\gamma_2=\gamma_2^\star$, the optimization problem $\max_{\gamma_1} \min_{i_1}  \min_{i_2} f_{i_1,i_2}(\gamma_1,\gamma_2^\star)$ satisfies \cite[Lemma 5]{arzou_arxiv} with $\gamma=\gamma_1$, $i=i_1$, and $k_i(\gamma)=\min_{i_2} f_{i,i_2}(\gamma,\gamma_2^\star)$. Hence, $\gamma_1^\star$ maximizing \eqref{2018.10.019_15.13_isit2019} satisfies
\begin{align}
\min_{i_2=1,2} f_{1,i_2}(\gamma_1^\star,\gamma_2^\star)=\min_{i_2=1,2} f_{2,i_2}(\gamma_1^\star,\gamma_2^\star),\label{isit2019_copy2_itw}
\end{align}
and in the case \eqref{isit2019_copy2_itw} does not have solution,
$\gamma_1^\star=0$ when $f_{1,i_2}(0,\gamma_2)> f_{2,i_2}(0,\gamma_2)$,  or $\gamma_1^\star=1$ otherwise. 
Combining \eqref{isit2019_copy1_itw} and \eqref{isit2019_copy2_itw} we obtain \eqref{optimal_gamma_isit2019}.

\section{Proof of Proposition \ref{LB_ISIT2019_bem}}
\label{proof_LB_ISIT2019_bem}
To prove Proposition \ref{LB_ISIT2019_bem}, 
we use the properties of the $\EsCR[\tau,i_1,i_2]{}(\cdot)$ function. Like always, $\nu\in\{1,2\}$, and $\nu^c$ denotes the complement index of $\nu$ among the set $\{1,2\}$.

 Let  $\gamma_\nu\in(\max_{u_\nu}P_{U_\nu}(u_\nu),1]$.
 In view of \eqref{B1_isit2019}, regardless of the value of $i_{\nu^c}$,
the minimization problem in the left hand side of \eqref{bef_6_isit2019} is done over an empty set when $i_\nu=1$, which leads to
$\EsCR[\tau,i_1,i_2]{}(\cdot)\big|_{i_\nu=1}=-\infty$.
For the case
$\gamma_1\in(\max_{u_1}P_{U_1}(u_1),1]$ and $\gamma_2\in(\max_{u_2}P_{U_2}(u_2),1]$,
if we have 
$i_1=i_2=2$, and considering  \eqref{B2_isit2019}, the problem becomes a minimization problem  without any constraint over distribution $\hat P_{\munderbar U}$, leading to 
$\EsCR[\tau,2,2]{}(\rho,P_{\munderbar U},\gamma_\nu)=E_{s,\tau}(\rho,P_{\munderbar U})$.

 Similarly,
 when
 $\gamma_1\in[0,\min_{u_1}P_{U_1}(u_1))$
 and
 $\gamma_2\in[0,\min_{u_2}P_{U_2}(u_2))$, 
  if
 $i_1=i_2=1$, \eqref{bef_6_isit2019} becomes a minimization problem without any constraint over distribution $\hat P_{\munderbar U}$ meaning that $\EsCR[\tau,1,1]{}(\rho,P_{\munderbar U},\gamma_\nu)=E_{s,\tau}(\rho,P_{\munderbar U})$.
While, regardless of the value of $i_{\nu^c}$,
for $\gamma_\nu\in[0,\min_{u_\nu}P_{U_\nu}(u_\nu))$,
if
  $i_\nu=2$, again 
 the minimization is done over an empty set leading to  $\EsCR[\tau,i_1,i_2]{}(\cdot)\big|_{i_\nu=2}=-\infty$.
  In our analysis, it suffices to consider $\gamma_\nu=0$ or $\gamma_\nu=1$ to represent the cases where $\EsCR[\tau,i_1,i_2]{}(\cdot)$ is infinity. Letting $\nu=1,2$, the same reasoning yields
 \setlength{\arraycolsep}{0.0em}
\begin{align}
&\EsCR[\tau,1,i_2]{}(\cdot)\big|_{\substack{\gamma_1=1}}=\EsCR[\tau,2,i_2]{}(\cdot)\big|_{\substack{\gamma_1=0}}=\EsCR[\tau,i_1,1]{}(\cdot)\big|_{\substack{\gamma_2=1}}=\EsCR[\tau,i_1,2]{}(\cdot)\big|_{\substack{\gamma_2=0}}=-\infty,&
\label{inf_7_isit2019}\\
&\EsCR[\tau,1,1]{}(\cdot)\big|_{\substack{\gamma_1=0, \gamma_2=0}}=\EsCR[\tau,1,2]{}(\cdot)\big|_{\substack{\gamma_1=0, \gamma_2=1}}=
\EsCR[\tau,2,1]{}(\cdot)\big|_{\substack{\gamma_1=1, \gamma_2=0}}=\EsCR[\tau,2,2]{}(\cdot)\big|_{\substack{\gamma_1=1, \gamma_2=1}}
=E_{s,\tau}(\rho,P_{\munderbar U}).& 
\label{inf_5_isit2019}
\end{align}

From \eqref{f_small_isit2019}, we conclude that 
for the cases given by \eqref{inf_7_isit2019} and \eqref{inf_5_isit2019}, 
the function
$f_{i_1,i_2} (\gamma_1,\gamma_2)$ is either infinity or is the Gallager exponent, i.e.,
\begin{align}
\min_{\tau\in\left\{
 \{1\},\{2\},\{1,2\}
 \right\}}\max_{\rho\in[0,1]}
E_0(\rho,\Qu[\tau]{i_\tau},W\Qu[\tau^c]{i_{\tau}^c})
-E_{s,\tau}(\rho,P_{\munderbar U}).\label{karet_miad_isit2019}
\end{align}
For example, when $\gamma_1,\gamma_2\in\{0,1\}$, the function 
$f_{i_1,i_2} (0,0)$ is equal to 
\eqref{karet_miad_isit2019} when $i_1=i_2=1$, and is infinity for the rest combinations of $i_1$ and $i_2$.

As a result, when $\gamma_1,\gamma_2\in\{0,1\}$, from
\eqref{inf_7_isit2019} and \eqref{inf_5_isit2019} we find that 
$f_{i_1,i_2}(\munderbar \gamma)$ is finite in only one case, and it is infinity for other combinations of $i_1$ and $i_2$, more specifically
\begin{align}
\min_{i_1,i_2=1,2}  f_{i_1,i_2} (0,0)=f_{1,1}(0,0), \quad
\min_{i_1,i_2=1,2}  f_{i_1,i_2} (0,1)=f_{1,2}(0,1), \label{nej1_isit2019} \\
\min_{i_1,i_2=1,2}  f_{i_1,i_2} (1,0)=f_{2,1}(1,0), \quad
\min_{i_1,i_2=1,2}  f_{i_1,i_2} (1,1)=f_{2,2}(1,1). \label{nej2_isit2019}
\end{align}

Next, by considering \eqref{nej1_isit2019} and \eqref{nej2_isit2019}, we lower bound 
the achievable exponent given by \eqref{2018.10.019_15.13_isit2019}. By taking maximization over 
$\gamma_\nu\in\{0,1\}$, rather than the interval of $[0,1]$, i.e., 
\begin{align}
E(P_{\munderbar U},W)
\geq 
\max_{\gamma_1,\gamma_2\in\{0,1\}}  \min_{i_1,i_2=1,2}  f_{i_1,i_2} (\gamma_1,\gamma_2)
,\label{LB_2018.10.019_15.13_isit2019}
\end{align}
we can find the following lower bound for $E(P_{\munderbar U},W)$
\begin{align}
E(P_{\munderbar U},W)
\geq \max \left\{
\min_{i_1,i_2=1,2}  f_{i_1,i_2} (0,0), 
\min_{i_1,i_2=1,2}  f_{i_1,i_2} (0,1),
\min_{i_1,i_2=1,2}  f_{i_1,i_2} (1,0),
\min_{i_1,i_2=1,2}  f_{i_1,i_2} (1,1)
\right\},\label{2018.12.12_17.28}
\end{align}
where by applying \eqref{nej1_isit2019} and \eqref{nej2_isit2019} 
into the minimizations over $i_1$ and $i_2$, 
we rewrite \eqref{2018.12.12_17.28} as
\begin{align}
E(P_{\munderbar U},W)
\geq \max \left\{
f_{1,1} (0,0), 
f_{1,2} (0,1),
 f_{2,1} (1,0),
f_{2,2} (1,1)
\right\}.\label{2018.12.12_17.29}
\end{align}
Inserting \eqref{karet_miad_isit2019} into 
\eqref{2018.12.12_17.29}, we conclude \eqref{lb_isit2019}.

\section{}
\label{ham_kare o hich kare}
In this appendix, we provide a number of general  lemmas that will be used
throughout the paper.
\begin{lemma}
\label{removing one minimization_noncondi} 
Let $T$ and $Z$ be two correlated random variables characterized by $P_{TZ}=P_{U}$.
For a given channel $W$, source $P_U=P_{TZ}$, and input distribution $Q$, 
let $E$ be 
\begin{align}
E=\min_{\hat{P}_{  U}\in \mathcal{P}_{\mathcal{ U}}}
\min_{\hat{P}_{  X Y}\in \mathcal{P}_{ \mathcal{ X}\times\mathcal{Y}}}
D(\hat P_{ U }|| P_{ U})+D(\hat P_{  X Y}||QW)
\hspace{12em}\nonumber\\
+\Big[\min_{\tilde{P}_{U}\in \mathcal{K}_s(\hat P_{ U })} \min_{\tilde{P}_{XY}\in \mathcal{K}_c(\hat P_{  X Y})} D(\tilde{P}_{  X Y}|| Q\hat{P}_{ Y})-H(\tilde P_{T|Z })\Big]^+,\label{the_weeping_meadow_2018}
\end{align}
where
\begin{align}
&\mathcal{K}_{s}(\hat P_{ U})\triangleq \left \{  \tilde{P}_{ U }
\in \mathcal{P}_{\mathcal{ U}} :
\tilde P_{ Z}  =\hat P_{ Z  },\ \mathbb{E}_{\tilde{P}}\log\big(P_{U}( U) \big) \geq \mathbb{E}_{\hat P}\log\big(P_{ U}( U) \big) \right \},
\label{k1n_21may_sou}
\\
&\mathcal{K}_{c}(\hat P_{  X Y})\triangleq \left \{  \tilde{P}_{  X Y}\in \mathcal{P}_{ \mathcal{ X}\times\mathcal{Y}} : 
\tilde P_{   Y}=\hat P_{  Y},\ \mathbb{E}_{\tilde{P}}\log\big(W(Y| X) \big) \geq \mathbb{E}_{\hat P}\log\big(W(Y| X) \big) \right \}.
\label{k1n_21may}
\end{align}
 It can be proved that
\begin{align}
E\geq
\min_{\hat{P}_{ U}\in \mathcal{P}_{\mathcal{ U}}}
\min_{\hat{P}_{  X Y}\in \mathcal{P}_{
\mathcal{ X}\times\mathcal{Y}}}
D(\hat P_{ U }|| P_{ U})+
D(\hat P_{ X Y}||  Q W)
+\Big[ D(\hat{P}_{  X Y}||Q\hat{P}_{ Y})-H(\hat P_{T|Z })\Big]^+.\label{12_27_21May}
\end{align}
\end{lemma}
\begin{IEEEproof}
Firstly,  assume for the optimal $ \hat P_U $, $\hat P_{XY}$,  $ \tilde P_U $ and $\tilde P_{XY}$ minimizing 
\eqref{the_weeping_meadow_2018}, we have
\begin{align}
D(\tilde{P}_{  X Y}|| Q\hat{P}_{ Y})-H(\tilde P_{T|Z })\geq 
D(\hat{P}_{  X Y}|| Q\hat{P}_{ Y})-H(\hat P_{T|Z }),\label{12_30_21May}
\end{align}
which leads to
\begin{align}
\Big[D(\tilde{P}_{  X Y}|| Q\hat{P}_{ Y})-H(\tilde P_{T|Z })\Big]^+\geq 
\Big[D(\hat{P}_{  X Y}|| Q\hat{P}_{ Y})-H(\hat P_{T|Z })\Big]^+.\label{12_31_21May}
\end{align}
Adding $D(\hat P_{ U }|| P_{ U})+D(\hat P_{  X Y}||QW)$ to the both sides of \eqref{12_31_21May}, \eqref{12_27_21May} is proved. 
Alternatively, if
\begin{align}
D(\tilde{P}_{  X Y}|| Q\hat{P}_{ Y})-H(\tilde P_{T|Z })\leq 
D(\hat{P}_{  X Y}|| Q\hat{P}_{ Y})-H(\hat P_{T|Z }),\label{12_40_21May_h}
\end{align} 
in view of
\eqref{k1n_21may_sou}, since $\tilde P_Z(z)=\hat P_Z(z)$, for all $z\in \mathcal Z$, we add $-H(\tilde P_Z)=-H(\hat P_Z)$ to the both sides of \eqref{12_40_21May_h}, where since $P_{T|Z}=\frac{P_U}{P_Z}$, we have
\begin{align}
D(\tilde{P}_{  X Y}|| Q\hat{P}_{ Y})+\sum_u \tilde P_U(u)\log\big(
\tilde P_U(u)
\big)\leq 
D(\hat{P}_{  X Y}|| Q\hat{P}_{ Y})
+\sum_u \hat P_U(u)\log\big(
\hat P_U(u)
\big)
.\label{12_40_21May}
\end{align}
Next, 
by using $\mathbb{E}_{\tilde{P}}\log\big(P_{ U}( U) \big) \geq \mathbb{E}_{\hat P}\log\big(P_{ U}( U) \big)$ and $\mathbb{E}_{\tilde{P}}\log\big(W(Y| X) \big) \geq \mathbb{E}_{\hat P}\log\big(W(Y| X) \big)$, respectively 
 given by \eqref{k1n_21may_sou} and \eqref{k1n_21may}, we find
\begin{align}
\sum_{ u}\tilde{P}_{U }( u )\log\big(P_{U}( u)\big)+
\sum_{  x,y}\tilde{P}_{ X Y}( x,y )\log\big(W(y| x)\big)\geq\hspace{10em}\nonumber\\
\sum_{ u}\hat{P}_{U }( u)\log\big(P_{ U}(u)\big)
+\sum_{ x,y}\hat{P}_{X Y}( x,y )\log\big(W(y| x)\big).\label{12_41_21May}
\end{align}
Subtracting \eqref{12_41_21May} from \eqref{12_40_21May} leads to
\begin{align}
\sum_{ u}  \tilde P_{U }(u)\log\bigg(\dfrac{\tilde P_{U}(u) }{P_{ U}( u)}\bigg)
+
\sum_{ x,y}  \tilde P_{  X Y }(  x,y)\log\bigg(\dfrac{ \tilde P_{  X Y
} (  x,y)}{W(y| x) Q(x)\hat P_{Y}(y)}\bigg)
\leq\hspace{5em} \nonumber\\
\sum_{u} \hat P_{ U  }(u)\log\bigg(\dfrac{\hat P_{U}(u)}{P_{ U}( u)}\bigg)
+
\sum_{x,y} \hat P_{  X Y }( x,y)\log\bigg(\dfrac{\hat P_{ XY} ( x,y)}{W(y| x)Q(x)\hat P_{Y}(y)}\bigg)
.\label{ned_eq_21May}
\end{align}
Moreover, in view of \eqref{k1n_21may}, $\tilde{P}_{ Y}=\hat{P}_{ Y}$ which yields  $H(\tilde{P}_{ Y})=H(\hat{P}_{ Y})$ or equivalently 
\begin{align}
\sum_{x, y}\tilde{P}_{  X Y}( x, y)
\log \hat{P}_{Y}( y)
= \sum_{ x, y}\hat{P}_{ X Y}( x, y)
\log \hat{P}_{ Y}( y).\label{h_tilde_hat_21may}
\end{align}
By adding \eqref{h_tilde_hat_21may} to the both sides of \eqref{ned_eq_21May}, we have
\begin{align}
\sum_{ u}  \tilde P_{U }(u)\log\bigg(\dfrac{\tilde P_{U}(u) }{P_{ U}( u)}\bigg)
+
\sum_{ x,y}  \tilde P_{  X Y }(  x,y)\log\bigg(\dfrac{ \tilde P_{  X Y
} (  x,y)}{W(y| x) Q(x)}\bigg)
\leq\hspace{5em} \nonumber\\
\sum_{u} \hat P_{ U  }(u)\log\bigg(\dfrac{\hat P_{U}(u)}{P_{ U}( u)}\bigg)
+
\sum_{x,y} \hat P_{  X Y }( x,y)\log\bigg(\dfrac{\hat P_{ XY} ( x,y)}{W(y| x)Q(x)}\bigg).\label{D_tilde_lower_hat_21may}
\end{align}
Using the definition of the relative entropy,
  \eqref{D_tilde_lower_hat_21may} can be expressed as
\begin{align}
D(\tilde{P}_{ U }||P_{ U})+
D(\tilde{P}_{  X Y}||Q W)
\leq D(\hat{P}_{ U }||P_{ U})+D(\hat{P}_{ X Y}||QW).\label{oz2_21may}
\end{align}
By adding $\Big[ D(\tilde{P}_{  X Y}||Q\hat{P}_{ Y})-H(\tilde P_{T|Z })\Big]^+$ on the both sides of \eqref{oz2_21may},
we obtain 
\begin{align}
D(\tilde{P}_{ U }||P_{ U})+
D(\tilde{P}_{  X Y}||Q W)
+\Big[ D(\tilde{P}_{  X Y}||Q\hat{P}_{ Y})-H(\tilde P_{T|Z })\Big]^+\leq
\hspace{8em}\nonumber\\
 D(\hat{P}_{ U }||P_{ U})+D(\hat{P}_{ X Y}||QW)
+
\Big[ D(\tilde{P}_{  X Y}||Q\hat{P}_{ Y})-H(\tilde P_{T|Z })\Big]^+.\label{oz}
\end{align}
Inasmuch as $\mathcal{K}_s(\hat P_{ U })\subset \mathcal P_{\mathcal U}$ and
$\mathcal{K}_c(\hat P_{  X Y})\subset \mathcal P_{ \mathcal X\times \mathcal Y}$, we have proved that whether 
$D(\tilde{P}_{  X Y}||Q\hat{P}_{ Y})-H(\tilde P_{T|Z })$ be lower than 
$D(\hat{P}_{  X Y}||Q\hat{P}_{ Y})-H(\hat P_{T|Z })$
or greater, we have \eqref{12_27_21May}.
\end{IEEEproof}
\begin{lemma}
\label{app_E0_0}
For a given channel $W$ with input distribution $Q$, we have
\begin{align}
\min_{\hat P_{XY}\in \mathcal{P}_{\mathcal{X}\times\mathcal{Y}}} D(\hat P_{XY}||Q W)+ \rho D(\hat P_{XY}||Q \hat P_Y)=E_0(\rho,Q,W),\label{2_min_m}
\end{align}
where $E_0(\rho,Q,W)=-\log\left(
\sum_{y}\left(\sum_x Q(x) W(y|x)^{\frac{1}{1+\rho}}
\right)^{1+\rho} \right)$. 
\end{lemma}
\begin{IEEEproof}
Firstly, we show that 
\begin{align}
D(\hat P_{XY}||Q\hat P_Y)=\min_{V_Y}D(\hat P_{XY}||QV_Y),\label{lema3_formul}
\end{align}
where $V_Y$ is an arbitrary probability assignment over the  alphabet  $\mathcal{Y}$. To prove \eqref{lema3_formul}, it suffices to show that $D(\hat P_{XY}||Q \hat P_Y)\leq D(\hat P_{XY}||Q V_Y)$ with equality if $\hat P_Y(y)=V_Y(y)$ for all $y\in\mathcal Y$. 
Subtracting $D(\hat P_{XY}||Q V_Y)$ from $D(\hat P_{XY}||Q \hat P_Y)$ leads to
\begin{align}
D(\hat P_{XY}||Q\hat P_Y)-D(\hat P_{XY}||Q V_Y)=\sum_{x,y}\hat P_{XY}(x,y)\log\dfrac{V_Y(y)}{\hat P_Y(y)}=-D(V_Y||\hat P_y)\leq 0, \label{41}
\end{align}
where \eqref{41} follows from the fact that the relative entropy is non-negative with equality when $V_Y(y)=\sum_x \hat P_{XY}(x,y)$ for all $y\in\mathcal{Y}$.
Thus,  \eqref{41}  yields
$
D(\hat P_{XY}||Q \hat P_Y)\leq D(\hat P_{XY}||Q V_Y)
$
and equality holds if $V_Y(y)=\hat P_Y(y)$ for all $y\in\mathcal{Y}$.
Hence, $D(\hat P_{XY}||Q \hat P_Y)
=\min_{V_Y}  D(\hat P_{XY}||Q V_Y)
$.

Next, by substituting \eqref{lema3_formul} into the left hand side of \eqref{2_min_m}, it remains to show that
\begin{align}
\min_{V_Y}\min_{\hat P_{XY}\in \mathcal{P}_{\mathcal{X}\times\mathcal{Y}}} D(\hat P_{XY}||Q W)+ \rho D(\hat P_{XY}||Q V_Y)=E_0(\rho,Q,W).\label{2_min}
\end{align}
In order to prove \eqref{2_min}, we start by applying Lagrange duality theory to the 
 inner minimization over $\hat P_{XY}$ in \eqref{2_min}. By recognizing  that 
the sum of the probabilities
of all possible outcomes must be $1$, the Lagrangian of optimization problem over 
$\hat P_{XY}$
 can be expressed as
\begin{align}
\Lambda(\hat P_{XY},\theta)=D(\hat P_{XY}||Q W)+ \rho D(\hat P_{XY}||Q V_Y) +\theta\Big(1- \sum_{x,y}\hat P_{XY}(x,y)\Big),
\end{align}
where  since the objective function  is convex with respect to $\hat P_{XY}$ and 
the constraint  $ \sum_{x,y}\hat P_{XY}(x,y)=1$ is affine, strong duality holds which leads to
\begin{align}
\min_{\hat P_{XY}\in \mathcal{P}_{\mathcal{X}\times\mathcal{Y}}} D(\hat P_{XY}||Q W)+ \rho D(\hat P_{XY}||Q V_Y)=\max_{\theta}\min_{\hat P_{XY}} \Lambda(\hat P_{XY},\theta).\label{kh0}
\end{align}

Using the definition of the relative entropy, the Lagrangian is simplified as
\begin{align}
\Lambda(\hat P_{XY},\theta)=\sum_{x,y}\hat P_{XY}(x,y) \log\dfrac{\hat P_{XY}(x,y)^{1+\rho}}{Q(x)^{1+\rho}W(y|x)V_Y(y)^\rho} +\theta\Big(1- \sum_{x,y}\hat P_{XY}(x,y)\Big).\label{kh1}
\end{align}
Since strong duality holds,  we can  proceed by analyzing the necessary KKT conditions. Setting
$\frac{\partial \Lambda(\hat P_{XY})}{\partial
 \hat{P}_{ X Y}(x,y)}=0$ yields
\begin{align}
\log\dfrac{\hat P_{XY}(x,y)^{1+\rho}}{Q(x)^{1+\rho}W(y|x)V_Y(y)^\rho}+(1+\rho)-\theta=0,
\end{align}
leading to
\begin{align}
\hat P_{XY}(x,y)=e^{\frac{\theta-(1+\rho)}{1+\rho}}Q(x)W(y|x)^{\frac{1}{1+\rho}}V_Y(y)^{\frac{\rho}{1+\rho}}.\label{kh2}
\end{align}
Summing both sides of \eqref{kh2} over $x,y$ and applying $\sum_{x,y} \hat P_{XY}(x,y)=1$, we obtain
\begin{align}
1=e^{\frac{\theta-(1+\rho)}{1+\rho}}\sum_{x,y}Q(x)W(y|x)^{\frac{1}{1+\rho}}V_Y(y)^{\frac{\rho}{1+\rho}}.\label{kh3}
\end{align}
Putting back $e^{\frac{\theta-(1+\rho)}{1+\rho}}$ obtained in \eqref{kh3} into \eqref{kh2}, the optimal $\hat P_{XY}$ is given by
\begin{align}
\hat P_{XY}(x,y)=\dfrac{Q(x)W(y|x)^{\frac{1}{1+\rho}}V_Y(y)^{\frac{\rho}{1+\rho}}}{\displaystyle\sum_{\bar x,\bar y}Q(\bar x)W(\bar y|\bar x)^{\frac{1}{1+\rho}}V_Y(\bar y)^{\frac{\rho}{1+\rho}}}.\label{kh4}
\end{align}
Substituting \eqref{kh4} into \eqref{kh1}, yields
\begin{align}
\max_\theta\min_{\hat P_{XY}}\Lambda(\hat P_{XY},\theta)=-(1+\rho)\log\left( \sum_{x,y}Q(x)W(y|x)^{\frac{1}{1+\rho}}V_Y(y)^{\frac{\rho}{1+\rho}}\right),\label{kh4.5}
\end{align}
where 
by putting back \eqref{kh4.5}  into \eqref{kh0}, \eqref{2_min} can be written as
\begin{align}
\min_{V_Y}\min_{\hat P_{XY}\in \mathcal{P}_{\mathcal{X}\times\mathcal{Y}}} D(\hat P_{XY}||Q W)+ \rho D(\hat P_{XY}||Q V_Y)\hspace{10em}\nonumber\\
=\min_{V_Y}-(1+\rho)\log\left( \sum_{x,y}Q(x)W(y|x)^{\frac{1}{1+\rho}}V_Y(y)^{\frac{\rho}{1+\rho}}\right).\label{kh5}
\end{align}

Since 
the  function in the log term of \eqref{kh5}  is a concave function with respect to $V_Y$ and the logarithm is an increasing function, \eqref{kh5} can be simplified as
\begin{align}
\min_{V_Y}\min_{\hat P_{XY}\in \mathcal{P}_{\mathcal{X}\times\mathcal{Y}}} D(\hat P_{XY}||Q W)+ \rho D(\hat P_{XY}||Q V_Y)\hspace{14em}\nonumber\\
=-(1+\rho)\log\left( \max_{V_Y} \sum_{x,y}Q(x)W(y|x)^{\frac{1}{1+\rho}}V_Y(y)^{\frac{\rho}{1+\rho}}\right),\label{59_13march}
\end{align}
where the optimization problem of the
right hand side of \eqref{59_13march} is solved by using  Lemma \ref{Vyyy}.

 Setting 
$e(y)=\sum_x Q(x)W(y|x)^{\frac{1}{1+\rho}}$ in Lemma \ref{Vyyy}, from \eqref{59_13march}  we obtain
\begin{align}
\min_{V_Y} \min_{\hat P_{XY}\in \mathcal{P}_{\mathcal{X}\times\mathcal{Y}}} D(\hat P_{XY}||Q W)+ \rho D(\hat P_{XY}||Q V_Y)\hspace{15em}\nonumber\\
=-(1+\rho)\log\left(\sum_y\bigg(\sum_x Q(x)W(y|x)^{\frac{1}{1+\rho}}\bigg)^{1+\rho}\right)^{\frac{1}{1+\rho}}.\label{kh9}
\end{align}
Applying the identity that $(1+\rho)\log(b^{\frac{1}{1+\rho}})=\log(b)$ to the right hand side of \eqref{kh9}, in view of the definition of $E_0(\cdot)$ and \eqref{2_min},
we conclude \eqref{2_min_m}.
\end{IEEEproof}
\begin{lemma}
\label{Vyyy}
Let $V_Y$ be a probability distribution and 
 $e(y)$ be a positive function such that for $\rho\in[0,1]$, the quantity $\sum_{y}e(y)V_Y(y)^{\frac{\rho}{1+\rho}}$ is a concave function of $V_Y(y)$. Then, we have
\begin{align}
\max_{V_Y} \sum_{y}e(y)V_Y(y)^{\frac{\rho}{1+\rho}}=\left( \sum_y e(y)^{1+\rho}\right)^{\frac{1}{1+\rho}},\label{Vy_13march}
\end{align}
where the optimal $V_Y$ maximizing \eqref{Vy_13march} is obtained as $
V_Y(y)=\dfrac{e(y)^{1+\rho}}{\sum_{\bar y} e(\bar y)^{1+\rho}}
$.
\end{lemma}
\begin{IEEEproof}
 Recalling that $\sum_y V_Y(y)=1$, the Lagrangian associated with the  optimization
problem in \eqref{Vy_13march} can be written  as
\begin{align}
\Lambda(V_Y,\theta)=\sum_{y}e(y)V_Y(y)^{\frac{\rho}{1+\rho}}+\theta(1-\sum_y V_Y(y)).
\end{align}
In view of the KKT condition, setting the partial derivative of $\Lambda(V_Y,\theta)$ with respect to $V_Y(y)$ equal to zero, yields 
\begin{align}
\dfrac{\rho}{1+\rho}e(y) V_Y(y)^{\frac{-1}{1+\rho}}-\theta=0.\label{gand_19_6_2018_18_42}
\end{align}
Solving \eqref{gand_19_6_2018_18_42} with respect to $V_Y(y)$ and applying the constraint that  $\sum_y V_Y(y)=1$,
 the optimal value of $V_Y(y)$ is derived as
$
V_Y(y)=\dfrac{e(y)^{1+\rho}}{\sum_{\bar y} e(\bar y)^{1+\rho}}
$.
Inserting the optimal $V_Y(y)$ into the left hand side of \eqref{Vy_13march}
proves Lemma \ref{Vyyy}.
\end{IEEEproof}
\bibliographystyle{IEEEtran}	
\bibliography{IEEEabrv,gcitation}

\end{document}